\documentclass[prd,preprint,superscriptaddress,nofootinbib,longbibliography,aps]{revtex4-1}
\usepackage[utf8]{inputenc}
\usepackage{graphicx}
\usepackage{amsmath}
\usepackage{amsfonts}
\usepackage{mathtools}
\usepackage{color}
\usepackage{amssymb}
\usepackage{dcolumn}
\usepackage{bm}
\usepackage{braket}
\usepackage{microtype} 
\usepackage[linktoc=all]{hyperref}
\usepackage{cleveref}
\usepackage{yfonts}
\usepackage{tensor}
\numberwithin{equation}{section}
\newcommand{\lag}{\mathcal{L}}
\newcommand{\ud}{\mathrm{d}}

\newcommand{\bel}[1] {\begin{equation}\label{#1}}
\newcommand{\beal}[1] {\begin{eqnarray}\label{#1}}

\newcommand{\be}{\begin{equation}}
\newcommand{\bea}{\begin{eqnarray}}
\newcommand{\ee}{\end{equation}}
\newcommand{\eea}{\end{eqnarray}}

\makeatletter
\DeclareRobustCommand{\rcite}[1]{%
\rcite@aux#1,\@nil{#1}%
}
\def\rcite@aux#1,#2\@nil#3{%
\if\relax#2\relax
Ref.~\cite{#3}%
\else
Refs.~\cite{#3}%
\fi
}
\makeatother

\begin{document}
\title{Shift symmetries, soft limits, and the double copy  \\beyond leading order}

\author{Mariana Carrillo Gonz\'alez} 
\email{cmariana@sas.upenn.edu}
\affiliation{Center for Particle Cosmology, Department of Physics and Astronomy,
	University of Pennsylvania, Philadelphia, Pennsylvania 19104, USA}
\author{Riccardo Penco}
\email{rpenco@cmu.edu}
\affiliation{Department of Physics, Carnegie Mellon University, Pittsburgh, PA 15213, USA}
\author{Mark Trodden}
\email{trodden@upenn.edu}
\affiliation{Center for Particle Cosmology, Department of Physics and Astronomy,
	University of Pennsylvania, Philadelphia, Pennsylvania 19104, USA}

\date{\today}

\begin{abstract}
\noindent 
In this paper, we compute the higher derivative amplitudes arising from shift symmetric-invariant actions for both the non-linear sigma model and the special galileon symmetries, and provide explicit expressions for their Lagrangians. We find that, beyond leading order, the equivalence between shift symmetries, enhanced single soft limits and compatibility with the double copy procedure breaks down.  In particular, we have shown that the most general even-point amplitudes of a colored-scalar satisfying the Kleiss-Kuijf (KK) and Bern-Carrasco-Johansson (BCJ) relations are compatible  with the non-linear sigma model symmetries. Similarly, their double copy is compatible with the special galileon symmetries. We showed this by fixing the dimensionless coefficients of these effective field theories in such a way that the arising amplitudes are compatible  with the  double copy procedure. We find that this can be achieved for the even-point amplitudes, but not for the odd ones. These results imply that not all operators invariant under the shift symmetries under consideration are compatible with the double copy.
\end{abstract}

\maketitle

\tableofcontents

\newpage

\section{Introduction}
In recent years, there has been a resurgent interest in exploring the infrared behavior of field theories and its implications (see {\it e.g.} \cite{Strominger:2017zoo} and references therein). While most of the attention has been devoted to gauge theories, interesting results have also been derived regarding the infrared structure of scalar effective field theories (see {\it e.g.}~\cite{Cheung:2014dqa,Cheung:2016drk,Low:2015ogb,Bogers:2018zeg,Yin:2018hht,Du:2015esa,Low:2014nga,Roest:2019oiw}.) For instance, Lorentz-invariant scalar field theories have been classified in~\cite{Cheung:2014dqa,Cheung:2016drk} according to their soft behavior and their numbers of derivatives per field. Among these, there are three interacting theories---the $U(N)$ non-linear sigma model (NLSM), the  Dirac-Born-Infeld (DBI) theory,  and the special galileon (SGal)~\cite{Hinterbichler:2015pqa,Cheung:2014dqa,Cachazo:2013hca,Cachazo:2014xea}---whose effective Lagrangians at lowest order in the derivative expansion each contain a single free parameter. These theories arise naturally in the Cachazo-He-Yuan (CHY) representation~\cite{Cachazo:2013hca,Cachazo:2014xea,Cachazo:2013gna,Cachazo:2013iea}, and are known collectively as ``exceptional scalar theories''.

Exceptional scalar theories display two noteworthy properties at leading order. First, their scattering amplitudes have an enhanced single soft limit, which follows from the invariance of the actions under non-linearly realized symmetries. Because of this feature, higher-point amplitudes can be obtained recursively from lower-point ones using a modified version of the Britto-Cachazo-Feng-Witten (BCFW) recursion relation~\cite{Britto:2005fq,Kampf:2012fn,Kampf:2013vha}. The second interesting property of exceptional scalar theories is that they are part of a web of theories related to each other by different implementations of color-kinematics replacements~\cite{Cachazo:2014xea,Cheung:2016prv,Cheung:2017ems,Cheung:2017yef,Zhou:2018wvn,Bollmann:2018edb}---see Figure 1 in~\cite{Carrillo-Gonzalez:2018pjk} for a pictorial  summary of these relations.

One of these color-kinematics relations is an especially relevant one that is known as the double copy. The double copy construction relates  colored-theories which satisfy the color-kinematics duality with their ``kinematic square'' \cite{Bern:2008qj,Bern:2010yg,Bern:2010ue} (for a pedagogical review see \cite{Elvang:2015rqa}.) The best-known version of this relation constructs gravitational\footnote{The gravitational theory which corresponds to the double copy of YM not only consists of a graviton, but also a dilaton and a 2-form field.} scattering amplitudes as  the double copy of  Yang-Mills (YM) scattering amplitudes. A similar double copy construction connects two of the exceptional scalar theories mentioned above giving rise to a relation that can be summarized as NLSM$^2$ = SGal. From this perspective, the NLSM and SGal can be thought of as scalar analogs of YM and gravity. In fact, the origin of such correspondence has been explored in different settings and can be understood as following from YM$^2$ = gravity after performing a ``dimensional reduction'' to extract the longitudinal modes~\cite{Cachazo:2014xea,Cheung:2017yef}. 
 Recently, it has been shown that the double copy holds not only for scattering amplitudes, but also for both exact and perturbative classical solutions~\cite{Monteiro:2014cda,Luna:2015paa,Luna:2016due,Ridgway:2015fdl,Carrillo-Gonzalez:2017iyj,Bahjat-Abbas:2017htu,Lee:2018gxc,Ilderton:2018lsf,Cho:2019ype,CarrilloGonzalez:2019gof,Luna:2018dpt,Berman:2018hwd,Saotome:2012vy,Neill:2013wsa,Luna:2016hge,Goldberger:2016iau,Goldberger:2017frp,Goldberger:2017vcg,Goldberger:2017ogt,Luna:2017dtq,Chester:2017vcz,Shen:2018ebu,Plefka:2018dpa,Kosower:2018adc,Maybee:2019jus}.

At lowest order in the derivative expansion, exceptional scalar theories can be equivalently defined through their symmetries, their enhanced single soft limits, and color-kinematics dualities. Importantly though, the inclusion of higher-order operators spoils this equivalence. For instance, it is clear that corrections with a large enough number of derivatives per field will not modify the soft limit, regardless of whether or not they preserve the symmetries. However, the status of color-kinematics duality is a priori less clear. In this paper, we focus on the NLSM$^2= $ SGal relation and explore the extent to which higher derivative corrections to these theories, consistent with their symmetries, are compatible with  color-kinematics duality. 

The analogous question has previously been asked for the YM$^2=$ gravity correspondence, and the higher-order operators of YM and their compatibility with the double copy has been explored in \cite{Broedel:2012rc,He:2016iqi,Cofano:2015jva}. While the $F^\mu{}_\nu F^\nu{}_\lambda F^\lambda{}_\mu$ term was shown to be compatible with the double copy, not all the $\mathcal{O}(F^4)$ contributions are compatible---not even the ones arising from the low energy limit of string theory. It is presently unknown whether there are hidden symmetries which only give rise to higher-order corrections that satisfy the color-kinematics duality.

Higher-order corrections to the NLSM amplitudes have been computed by several different methods. These constructions do not rely on the symmetries of the NLSM but instead focus on satisfying the color-kinematics duality or on the infrared behavior of the theory. One construction, \cite{Carrasco:2016ldy}, consists of a rewriting of the open string amplitude in terms of a function called the Z-function involved in a Kawai-Lewellen-Tye (KLT)-like product with the YM amplitude. The Z-function behaves as a doubly-ordered partial amplitude and satisfies the Kleiss-Kuijf (KK)~\cite{Kleiss:1988ne} and Bern-Carrasco-Johansson (BCJ)~\cite{Bern:2008qj} relations. By taking the Abelian and $\alpha'\rightarrow 0$ limits, the Z-function reduces to the NLSM partial amplitudes.\footnote{We should emphasize that this result does not imply that the string spectrum includes NLSM scalars, but rather that information about their tree-level scattering amplitudes is hidden in the open string tree-level scattering amplitudes.} Given this, it has been proposed that the $\alpha'$-corrections correspond to the higher-order corrections to the NLSM. It is interesting to note that all odd-point amplitudes arising from this construction vanish. The theory giving rise to these amplitudes has been dubbed the Abelian~Z-theory.   A second construction \cite{Elvang:2018dco} starts from the most general color-ordered scalar 4-point amplitude up to 8th order in derivatives and imposes cyclicity, the KK relations, and the BCJ relations; all these requirements are highly constraining and completely fix the scattering amplitude. In fact, this 4-point amplitude coincides with that of the Abelian Z-theory. The authors of~\cite{Elvang:2018dco} also considered the 5-point amplitude, and showed that, while the contribution coming from the NLSM Wess-Zumino term does not satisfy the BCJ relations, there is a contribution at 14th order in derivatives that is compatible with the double copy prescription. Similarly, the 6-point function was computed up to 6th order in derivatives. A third method~\cite{Rodina:2018pcb}, assumes the pion double soft theorems~\cite{Du:2015esa,Low:2015ogb,Cachazo:2015ksa} to compute the higher-order corrections, and finds the same results as the Abelian Z-theory plus an additional correction to the 4-point amplitude at order $\mathcal{O}(p^4)$ which does not obey the BCJ relations. Earlier work along this lines was previously performed in \cite{Susskind:1970gf,Ellis:1970nn,Volkov:1973up}, and more recently in \cite{Cheung:2014dqa,Cheung:2015ota,Cheung:2016drk,Elvang:2018dco} . This method has now been dubbed the {\it soft bootstrap}. The soft bootstrap consists of constructing a modified BCFW recursion relation for scattering amplitudes based on the degree $\sigma$ of its soft theorem, which is defined by
\begin{equation}
\mathcal{A}_n(p_1,\dots,p_{n-1},\epsilon  p_n)\quad\stackrel{\epsilon\rightarrow 0}{\longrightarrow} \quad\epsilon^\sigma S_n+\cdots  \ ,
\end{equation}
with $S_n\neq 0$ a ``soft factor'' involving the first $n-1$ momenta.  Recently, it was shown that the soft bootstrap approach can be extended to $\mathcal{O}(p^4)$ for the NLSM \cite{Low:2019ynd}. In \cite{Low:2019ynd}, higher-point amplitudes at $\mathcal{O}(p^4)$ were obtained by defining soft blocks for 4 and 5 pions and using these as seeds in the soft bootstrap. As well as single-trace amplitudes, multi-trace amplitudes were also constructed, and both the $SU(N)$ and $SO(N)$ NLSMs were considered. Nevertheless, the extension to $\mathcal{O}(p^6)$ and higher is not completely  obvious.  Lastly, another way of obtaining the higher derivative corrections to the NLSM is through the ``extraction" of the longitudinal modes of YM, {\it i.e.} using the techniques of  \cite{Cheung:2017yef,Cheung:2017ems}. This was done in \cite{Mizera:2018jbh}, where the leading order Lagrangian of the Abelian Z-theory was found from a dimensional reduction of the $F^\mu{}_\nu F^\nu{}_\lambda F^\lambda{}_\mu$ YM term. 

Higher-order corrections to the SGal amplitudes have previously been considered in the literature, for instance in~\cite{Elvang:2018dco} by using the soft bootstrap.  Using this method, one can compute the  higher derivative  corrections to a theory from the leading order amplitudes by assuming the single soft limit. It has been shown that the  special galileon is the only interacting theory satisfying the soft limit with $\sigma=3$ non-trivially \cite{Hinterbichler:2014cwa,Bogers:2018zeg,Roest:2019oiw}. This limit is not only satisfied  non-trivially by its leading order amplitude, but also by several higher derivative corrections. It is important to note that not all higher-order amplitudes can be constructed using the soft bootstrap approach. This limitation  follows from the fact that the single soft limit can be trivially satisfied at sufficiently high order; for a term in the Lagrangian of the form $\partial^m\phi^n$, the soft limit of degree $\sigma$ becomes trivial if $\sigma\leq m/n$. Similarly, one should notice that satisfying a single soft limit does not imply that the amplitude comes from a shift symmetric theory\footnote{One could check if an amplitude comes from a shift symmetric theory by looking at the double soft theorem, which contains information about the algebra that is non-linearly realized. Note that the softbootstrap approach in \cite{Elvang:2018dco} only considers single soft limits. Other approaches have considered double soft limits \cite{Rodina:2018pcb}, but their applications are more restrictive than that of \cite{Elvang:2018dco}.}. For example, a term such as $(\partial\partial\partial\pi)^4$ would lead to an amplitude with soft degree $\sigma=3$, but it is not invariant under the special galileon symmetries. We should also note that the corrections computed by using the soft bootstrap method include a non-vanishing 5-point amplitude. A second approach consists of finding the special galileon corrections as the double copy of the NLSM corrections. By considering the double copy of the Abelian Z-theory one obtains the even-point special galileon higher-order amplitudes from \cite{Elvang:2018dco}. Finally, a third approach towards computing the higher derivative operators invariant under the special galileon symmetry was followed in \cite{Novotny:2016jkh}; the invariant Lagrangian was constructed up to quartic order in the Galileon field through a brane construction similar in spirit to \cite{deRham:2010eu,Goon:2011qf}.

\begin{figure}[t]
	\begin{center}
		\vspace{-2cm}
		\includegraphics[scale=0.44]{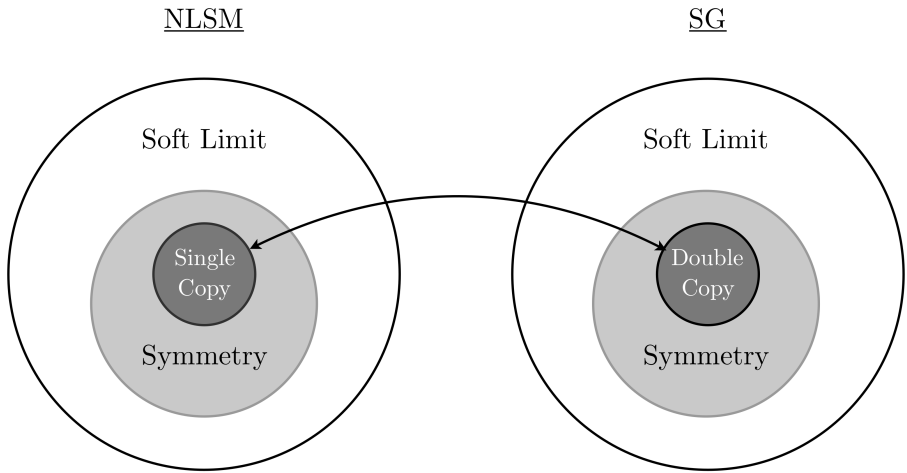}
		\vspace{-2cm}
		\caption{Diagram showing the relations between the even-point scattering amplitudes for different definitions of the NLSM and the SGal. The even-point amplitudes that are compatible with the double copy are also compatible with the NLSM and SGal symmetries. With respect to the odd-point amplitudes, there is a 5-point amplitude at the 14th derivative order which satisfies the KK and BCJ relations. We have shown that this amplitude  cannot arise from NLSM Lagrangian involving a Levi-Civita.  Nevertheless, it could be possible that it is compatible with the NLSM symmetries. The double copy of this term leads to a 5-point amplitude at the 32nd derivative order which could arise from the Special Galileon but is not considered here.} \label{fig:summary}
	\end{center}
\end{figure}

From these results, it is clear that the definitions of the exceptional scalar theories through their enhanced single soft limits, through their symmetries, or through the double copy, are only equivalent at leading order, and that this equivalence breaks down when including higher-order operators. In this paper, we will explore the definition of these theories as given by their shift symmetries. We will  not only  compute the  on-shell  scattering amplitudes, but we  will  find the shift  symmetric  Lagrangians giving rise  to them. The Lagrangian is relevant for calculations such  as the  classical  perturbative double copy in \cite{Carrillo-Gonzalez:2018pjk}. We will rely on a coset construction~\cite{Callan:1969sn, Volkov:1973vd, ogievetsky:1974ab} to write down the most general higher derivative corrections that are compatible with the SGal and NLSM symmetries. We will then constrain the NLSM coupling constants by demanding that the on-shell scattering amplitudes  satisfy the KK and BCJ relations in order to be able to construct the double copy. Here, we follow the approach of \cite{Broedel:2012rc} and assume that the double copy for higher order operators follows in the same way as it does for the leading order ones. Our goal is to understand whether the double copy of the higher-order corrections to the NLSM obtained this way corresponds to (a subset of) all possible higher-order corrections to the SGal theory. A pictorial summary of our results is provided in Figure \ref{fig:summary}. 

The rest of this paper is organized as follows. In Sec. \ref{coset}, we give a short review of the coset construction which will be used to build the higher derivative corrections to the NLSM and the SGal. In Sec. \ref{nlsm}, we analyze the higher derivative corrections to the $SU(N) \times SU(N) \to SU(N)$ NLSM in the large $N$ limit, and in Sec. \ref{sg} we explicitly construct the higher-order Lagrangian of the SGal. In  Sec. \ref{sec: double copy} we explore the extent to which the higher derivative corrections introduced in the previous two sections are compatible with color-kinematics duality. Finally, we discuss our results and conclude in Sec. \ref{conc}.

\section{Short review of the coset construction} \label{coset}

We begin by giving a brief review of the coset construction~\cite{Callan:1969sn} for spontaneously broken space-time symmetries~\cite{Volkov:1973vd,ogievetsky:1974ab}.  This construction is a method that allows the systematic construction of an effective field theory Lagrangian for Goldstone modes solely based on the knowledge of the symmetry breaking pattern. For recent, more detailed discussions see also~\cite{Goon:2012dy,Goon:2014ika,Delacretaz:2014oxa,Goon:2014paa}.

Consider a system whose ground state spontaneously breaks a symmetry group $G$, which contains the Poincar\'e group as a subgroup, down to a subgroup $H$. In general, $H$ may correspond to internal, space-time, or a mixture of both types of symmetries. We will denote the broken generators by $X_\alpha$, the unbroken translations by $P_a$, and the remaining unbroken symmetry generators by $T_A$. The effective action for the Goldstone bosons realizes both the unbroken translations and the broken symmetries non-linearly, while the other unbroken symmetries are implemented linearly and are therefore manifest.

The starting point of a coset construction is a dramatization of the most general symmetry transformation that is generated by the broken generators together with an unbroken translation:\footnote{Throughout this paper we will work with anti-hermitian generators. This will simplify the following equations by eliminating many factors of ``$i$''.} 
\begin{equation} \label{coset parametrization}
\Omega(x,\pi)=e^{x^a P_a}e^{\pi^\alpha X_\alpha} \ .
\end{equation}
Since $\Omega$ is defined only up to an overall unbroken symmetry transformation it is an element of a coset, hence the name of this construction. From this, one can define the Maurer-Cartan form
\begin{equation}\label{MC form}
\Omega^{-1}\ud\Omega =  \omega^a P_a+\omega^\alpha  X_\alpha+\omega^A T_A .
\end{equation}
This is an element of the algebra, and as such it can be written as a linear combination of all the generators. The coefficients of this expansion can be calculated explicitly using the algebra of $G$, the Baker-Campbell-Hausdorff formula, and various identities involving matrix exponentials.  The coefficients can be conveniently parametrized as follows:
\begin{subequations}
\begin{eqnarray}
	\omega^a &=& E^a{}_\mu \ud x^\mu \\
	\omega^\alpha &=& \nabla_a\pi^\alpha  E^a{}_\mu \ud x ^\mu \\
	\omega^B &=& A_a^B E^a{}_\mu \ud x ^\mu .
\end{eqnarray}
\end{subequations}

It can be shown~\cite{ogievetsky:1974ab} that the components $E^a_\mu$ play the role of a vielbein, in the sense that the volume element $\det(E) \,\ud^d x$ is a scalar under $G$. One can also check that the quantities $\nabla_a\pi^\alpha$---usually referred to as ``covariant derivatives'' of the Goldstone modes---transform under $G$ as a (possibly reducible) linear representation of $H$. Thus, contractions of such covariant derivatives that are manifestly invariant under $H$ are also secretly invariant under the full group $G$. Finally, the quantities $A_a^B$ transform as the components of a connection, and can be used to introduce a covariant derivative as follows:
\begin{equation} \label{coset cov dev}
\nabla_a\equiv{(E^{-1})_a}^\mu\partial_\mu+ A_a^B T_B \ .
\end{equation}
This definition allows us to calculate higher-order covariant derivatives of the Goldstones or, for that matter, covariant derivatives of any field that is charged under $H$.

We can now used the building blocks introduced above to write down the most general effective action for the Goldstone modes, which schematically takes the following form:
\begin{equation}
S = \int d^d x \det(E) f (\nabla_a\pi^\alpha,\nabla_b\nabla_a\pi^\alpha,\cdots) \ ,
\end{equation}
where all the indices are contracted in such a way as to preserve the unbroken symmetries. 

If only internal symmetries are broken, the number of Goldstone modes is equal to the number of broken generators---this is the usual Nambu-Goldstone theorem~\cite{Nambu:1960tm,Goldstone:1961eq}. However, when some of the symmetries that are spontaneously broken are space-time symmetries, one can usually obtain a non-linear realization of the symmetries that involves fewer fields~\cite{Low:2001bw}. Specifically, if commutation with some unbroken translation $P$ relates two multiplets (under $H$) $X$ and $X'$ of broken generators, {\it i.e.}
\begin{equation}
[P,X']\supset X, 
\end{equation}
then one can eliminate the Goldstones that would be naively associated with $X'$ and express them in terms of Goldstones of $X$ and their derivatives. This is done by imposing a set of so-called ``inverse Higgs constraints"~\cite{Ivanov:1975zq}, which amount to setting to zero (a subset of) covariant derivatives of the Goldstones of $X$ in the same representation as the Goldstones of $X'$. Given the transformation properties of the Goldstone covariant derivatives, this procedure can be shown to preserve all the symmetries---including the ones that are non-linearly realized.


\section{Higher-order Lagrangian for the non-linear sigma model} \label{nlsm}

In this section, we will consider a NLSM corresponding to the symmetry breaking pattern $G_L \times G_R \rightarrow G_\text{diag}$, where $G_L= G_R= G_\text{diag} \equiv G$ is a simple, compact, and internal symmetry group. For simplicity we will also restrict our attention to $d=4$ spacetime dimensions. We will first derive the main building blocks of the effective Lagrangian using a coset construction, and discuss two different choices of coset parameterizations.  Then, we will focus on the particular case where $G = SU(N)$, and write down all possible higher derivative corrections up to $\mathcal{O} (\partial^8)$ in the large-$N$ limit. In this limit, our results will also apply to $G = U(N)$.

\subsection{Coset construction and lowest-order effective Lagrangian}

Let us choose the broken generators $X_\alpha$ that appear in the coset parametrization \eqref{coset parametrization} to be the generators of, say, $G_L$. Then the components of the Maurer-Cartan form in \eqref{coset parametrization} read 
\begin{equation} \label{MC form NLSM}
	\Omega^{-1} \partial_\mu \Omega =  P_\mu + 2 f^{\alpha\beta\gamma} (U^{-1} \partial_\mu U)_{\beta \gamma} X_\alpha .
\end{equation}
where $f^{\alpha\beta\gamma}$ are the structure constants of the group $G$, and $U_{\alpha \beta}$ is the adjoint representation of the abstract group element $e^{\pi^\alpha X_\alpha}$. To derive the result above, we used the fact that, in the adjoint representation, the generators $X_\alpha$ are normalized as:
\begin{equation} \label{adjoint normalization}
	\text{Tr}\left(X_\alpha X_\beta\right)= - \tfrac{1}{2}\delta_{\alpha\beta} \, , \qquad \qquad \qquad \text{(adjoint)} \, .
\end{equation}
Note, however, that Eq. \eqref{MC form NLSM} follows exclusively from the algebra of the group and the symmetry breaking pattern, and it is valid in any representation.

The coset vielbein is trivial because the broken generators are all internal. Therefore, the covariant derivatives of the Goldstones $\pi^\alpha$ are simply
\begin{equation} \label{NLSM cov dev 1}
	\nabla_\mu \pi^\alpha = 2 f^{\alpha\beta\gamma} (U^{-1} \partial_\mu U)_{\beta \gamma} \ .
\end{equation}
Moreover, the Maurer-Cartan form does not have components along the unbroken generators, and therefore the coset covariant derivatives defined in \eqref{coset cov dev} reduce to ordinary partial derivatives, {\it i.e.} $\nabla_\mu = \partial_\mu$.

Because the commutators of the broken generators $X_\alpha$ with the unbroken generators $T_A$ of $G_\text{diag}$ read
\begin{equation}
	[ T_A, X_\beta ] = - f_{A \beta \gamma} X^\gamma ,
\end{equation}
the covariant derivatives $\nabla_\mu \pi^\alpha$ transform in the adjoint representation under $G_\text{diag}$. The effective Lagrangian must be manifestly invariant under all unbroken symmetries, and therefore up to quadratic order in derivatives  it must be\footnote{We are working with a metric with ``mostly minus'' signature.}
\begin{equation} \label{NLSM O(p2)}
	\mathcal L_\text{NLSM}^{(2)} = \frac{F^2}{8} \, \nabla_\mu \pi^\alpha \nabla^\mu \pi_\alpha , 
\end{equation}
where $F$ is the symmetry breaking scale, and the factor of $1/8$ has been added for later convenience. At lowest order in the Goldstones, the covariant derivatives are equal to ordinary derivatives, {\it i.e.} $\nabla_\mu \pi^\alpha \simeq \partial_\mu \pi^\alpha + \mathcal{O}(\pi \partial \pi)$, and thus, the canonically normalized fields are $\phi^\alpha \equiv F \pi^\alpha / 2 $. Higher derivative corrections to the Lagrangian \eqref{NLSM O(p2)} contain either higher powers of $\nabla_\mu \pi^\alpha$, or additional {\it ordinary} derivatives (as opposed to {\it covariant}, because the coset connection in \eqref{MC form NLSM} vanishes).

One of the advantages of the coset construction is that it does not rely on a specific representation of $G_L \times G_R$. This makes it explicit that the dynamics of the Goldstone modes depends solely on the symmetry breaking pattern, and not on the particular representation of the order parameter that realizes it. However, it can be instructive to rewrite the lowest order Lagrangian \eqref{NLSM O(p2)} that we obtained from the coset construction by assuming a particular representation. This will allow us to recast our result in a form that the reader might be more familiar with. 

To this end, we notice that Eqs. \eqref{MC form NLSM} and \eqref{NLSM cov dev 1} the following identity must be valid in any representation:
\begin{equation} \label{NLSM cov dev arbitrary rep}
	(\mathcal{U}^{-1} \partial_\mu \mathcal{U})_{IJ} = \nabla_\mu \pi^\alpha (X_\alpha)_{IJ} \, .
\end{equation}
with $\mathcal{U}_{IJ} \equiv (e^{\pi^\alpha X_\alpha})_{IJ}$.
In an arbitrary representation of an arbitrary group, the $X_\alpha$'s are normalized according to 
\begin{equation} \label{NLSM normalization arbitrary rep}
	\text{Tr}\left(X_\alpha X_\beta\right)= - \mathcal{T} \delta_{\alpha\beta},
\end{equation}
where $\mathcal{T}$ is the index of the representation. For instance, the indices of the fundamental representations of $SU(N)$ and $SO(N)$ are respectively equal to $1/2$ and $2$~\cite{Srednicki:2007ab}. Using the result \eqref{NLSM cov dev arbitrary rep} together with the normalization condition \eqref{NLSM normalization arbitrary rep}, it is easy to show that the lowest order Lagrangian \eqref{NLSM O(p2)} can be rewritten as
\begin{equation} \label{lowest order NLSM Lagrangian U}
	\mathcal L_\text{NLSM}^{(2)} =  \frac{F^2}{8 \mathcal{T}} \, \text{Tr} \, (\partial_\mu \mathcal{U}^{-1} \partial^\mu \mathcal{U}) \, .
\end{equation}
In the particular case of the fundamental representation of $G = SU(N)$, this reduces to the standard expression for the lowest order Lagrangian in chiral perturbation theory~\cite{Donoghue:1992dd}.

\subsection{Alternative coset parametrization}
\label{AltCos}

The explicit form of the covariant derivatives \eqref{NLSM cov dev 1} crucially relies on our choice that the broken generators $X_\alpha$ entering the coset parametrization \eqref{coset parametrization}  be the generators of $G_L$. This choice is convenient because in this case the $X_\alpha$'s span a subgroup, but of course it is not the only possible one, since the notion of broken generator is always determined only up to the addition of unbroken generators. Different choices for the $X_\alpha$'s lead to coset parametrizations which are related to each other by field redefinitions of the Goldstones. 

Another natural choice for the broken generators is $X_\alpha = \frac{1}{\sqrt{2}} (J^L_\alpha - J_\alpha^R) $, where $J_\alpha^{L,R}$ are the generators of $G_{L,R}$. It is easy to see that the components of the Maurer-Cartan form in this case read
\begin{equation}
	\Omega^{-1} \partial_\mu \Omega = P_\mu + \sqrt{2} f^{\alpha\beta\gamma} (U^{-1} \partial_\mu U)_{\beta \gamma} J^L_\alpha + \sqrt{2} f^{\alpha\beta\gamma} (U \partial_\mu U^{-1})_{\beta \gamma} J^R_\alpha . 
\end{equation}
By rewriting the right-hand side of this equation in terms of broken ($X_\alpha$) and unbroken ($T_\alpha =  \frac{1}{\sqrt{2}} (J^L_\alpha + J_\alpha^R)$) generators, we can read off the coset covariant derivatives and connections in this new parametrization: 
\begin{subequations}
\begin{align}
	\nabla_\mu \pi^\alpha &=  f^{\alpha\beta\gamma} (U^{-1} \partial_\mu U-U\partial_\mu U^{-1})_{\beta \gamma} ,\\
	A_\mu^\alpha &=  f^{\alpha\beta\gamma} (U^{-1} \partial_\mu U+U\partial_\mu U^{-1})_{\beta \gamma} . \label{conn}
\end{align}
\end{subequations}
The effective Lagrangian at lowest order in the derivative expansion is still \eqref{NLSM O(p2)}, but now with a slightly different expression for $\nabla_\mu \pi^\alpha$. Higher derivative corrections involve higher powers of $\nabla_\mu \pi^\alpha$, or covariant derivatives $\nabla_\mu f  \equiv \partial_\mu f + [A_\mu^\alpha \, T_\alpha,  f]$. In what follows, we will use this alternative coset parametrization to write down all non-redundant contributions to the NLSM effective Lagrangian up to eighth order in derivatives. This will enable us to leverage results that have already been derived in the context of chiral perturbation theory~\cite{Gasser:1984gg,Fearing:1994ga,Bijnens:1999sh,Ebertshauser:2001nj,Bijnens:2001bb,Bijnens:2018lez}.

\subsection{Higher-derivative corrections for $G = SU(N)$} \label{sec: higher der NLSM}

We will now specialize our analysis to the case where $G=SU(N)$ and work in the large-$N$ limit. This will allow us to  focus directly on those terms that are relevant for the double copy construction---see Sec.~\ref{Sec: NLSM double copy} for more details---and as an added bonus will also reduce the overall number of terms we need to include in the Lagrangian. Moreover, we will omit redundant terms that can be eliminated by a field redefinition (because these are proportional to the lowest order equations of motion), by performing integrations by parts, or by using the Bianchi and Levi-Civita identities summarized in Appendix \ref{App Bianchi Levi Civita}.

Another property that can be used to simplify the Lagrangian after expanding in powers of the Goldstone fields is the $SU(N)$ completeness relation 
\begin{equation}
{(J^\alpha)_{I}}^J {(J_\alpha)_{K}}^L= \frac{1}{2} \left( \frac{1}{N}{\delta_{I}}^J{\delta_{K}}^L - {\delta_{I}}^L{\delta_{K}}^J \right) \ ,
\end{equation}
where the first term, which would not be present for $G=U(N)$, leads to terms that are subleading in the large-$N$ limit. For particular values of $N$ there exist additional trace relations that can further reduce the basis of operators in the Lagrangian, but since we are interested in results that have more general validity we will not employ these here.

In order to make our notation a little more more compact, we will work with a particular representation of $SU(N)$---the fundamental representation---and we will define the quantity
\begin{equation}
	(u_\mu)_{IJ} \equiv  \nabla_\mu \pi^\alpha (X_\alpha)_{IJ} \, .
\end{equation}
We can then express the lowest order effective Lagrangian \eqref{NLSM O(p2)} directly in terms of $u_\mu$ as follows:
\begin{equation} \label{lNLSM}
	\mathcal L_\text{NLSM}^{(2)} = -\frac{F^2}{4} \, \text{Tr} \, (u_\mu u^\mu) .
\end{equation}

Once again, the canonically normalized field is $\phi^\alpha = F \pi^\alpha / 2$.

The next-to-leading order correction to this Lagrangian contains four derivatives and an arbitrary number of Goldstone fields, and reads~\cite{Gasser:1984gg}
\begin{align} 
\lag_\text{NLSM}^{(4)}&=c_1  \text{Tr}\left( u^\mu u^\nu u_\mu u_\nu\right) +c_2 \text{Tr}\left( u_\mu u^\mu u_\nu u^\nu\right) + \cdots \ , \label{lNLSM4}
\end{align}
where the $c_i$'s are constant dimensionless coefficients, and the ellipsis represents terms with more than one trace, which are negligible in the large $N$ limit~\cite{Kaiser:2000gs}. In the particular case of $N=3$, the  first term is redundant and can be expressed as a combination of the second one with terms involving more than one trace~\cite{Donoghue:1992dd}; for $N=2$ the second term is also redundant, and therefore all terms with four derivatives can be written as multi-trace terms~\cite{Donoghue:1992dd}. 

At fourth order in derivatives, there is an additional single-trace term that can be added to the Lagrangian. This is the Wess-Zumino-Witten (WZW) term \cite{Wess:1971yu,Witten:1983tw}, and unlike the terms in \eqref{lNLSM4} it is invariant under $G$ only up to a total derivative. This term can be built by extending the base manifold to 5 dimensions, and introducing the invariant, exact 5-form
\begin{equation}
	d \beta \equiv \text{Tr}(u_\mu u_\nu u_\lambda u_\rho u_\sigma) \, \ud x^\mu \wedge \ud x^\nu \wedge \ud x^\lambda \wedge \ud x^\rho \wedge \ud x^\sigma \, . 
\end{equation}
Up to an overall coefficient, the integral of the 4-form $\beta$ over the space-time manifold is the WZW term. It is the only 4-derivative term in the Lagrangian that gives rise to odd-point functions. For instance, at leading order in an expansion in canonically normalized Goldstone fields it reads
\begin{equation} \label{NLSM WZW}
	\mathcal{L}_\text{WZW} \simeq \frac{c}{F^5} \, \epsilon^{\mu\nu\lambda\rho} \phi^\alpha\partial_\mu\phi^\beta\partial_\nu\phi^\gamma\partial_\lambda\phi^\delta\partial_\rho\phi^\varepsilon\, \text{Tr}(X_\alpha X_\beta X_\gamma X_\delta X_\varepsilon) . 
\end{equation}
Notice that the WZW term vanishes for $N=2$, whereas for $N=3$ the coefficient $c$ is famously quantized~\cite{Witten:1983tw}. Moreover, this term breaks the $\mathbb{Z}_2$ symmetry $\phi \to -\phi$, also known as {\it intrinsic parity}. Thus, this term (and others) can in principle be omitted, if desired, by requiring that such a symmetry be preserved. 

The 6-derivative corrections to our NLSM Lagrangian read~\cite{Fearing:1994ga,Bijnens:1999sh,Ebertshauser:2001nj,Bijnens:2001bb} 
\begin{align}
\lag_\text{NLSM}^{(6)} &=\frac{1}{F^2}\Big\{d_1\text{Tr}\left(u\cdot u \,h_{\mu\nu}h^{\mu\nu}\right)+d_2  \text{Tr}\left(h_{\mu\nu}u_\rho h^{\mu\nu}  u^\rho \right) +d_3   \text{Tr}\left(h_{\mu\nu}u_\rho h^{\mu\rho}  u^\nu \right)\nonumber \\
&+ e_1 \text{Tr}(h_{\mu\nu}u^\nu  u^\beta u^\gamma u^\delta)\epsilon_{\mu\beta\gamma\delta} +e_2 \text{Tr}(\nabla_\mu u^\nu\nabla_\nu u^\rho \nabla_\rho u^\mu)+f_1 \text{Tr}\left[(u\cdot u)^3\right]\nonumber \\&+f_2\text{Tr}\left(u\cdot u\,u_\mu\,u\cdot u\, \, u^\mu\right) +f_3\text{Tr}\left(u\cdot u\,u_\mu u_\nu u^\mu  u^\nu\right)+f_4\text{Tr}\left(u_\mu u_\nu u_\rho u^\mu u^\nu u^\rho\right)\nonumber \\&+f_5\text{Tr}\left(u_\mu u_\nu u_\rho u^\mu u^\rho u^\nu\right)+\cdots\Big\} \ , \label{lNLSM6}
\end{align}
where $h_{\mu\nu}=\nabla_\mu  u_\nu+\nabla_\nu u_  \mu  \ , $ and the ellipsis again denotes multi-trace contributions which are negligible in the large-$N$ limit. Moreover, the terms proportional to the $e_i$ coefficients break intrinsic parity, just like the Wess-Zumino term does, and give rise to odd-point amplitudes.  

Finally, at 8th order in the derivative expansion we have~\cite{Bijnens:2018lez} 
\begin{align}
\lag^{(8)}_\text{NLSM}&=\frac{1}{F^4}\Big\{g_1 \text{Tr}\left(\nabla^\mu u^\nu\nabla_\nu u^\rho\nabla_\rho  u^\sigma\nabla_\sigma u_\mu\right)+g_2 \text{Tr}\left(\nabla^\mu u^\nu\nabla^\rho u_\mu \nabla^\sigma  u_\rho\nabla_\nu u_\sigma\right) \nonumber  \\
&+g_3 \text{Tr}\left(\nabla^\mu u^\nu\nabla^\rho u_\nu \nabla_\mu  u^\sigma\nabla_\rho u_\sigma\right)+g_4 \text{Tr}\left(\nabla^\mu u^\nu\nabla^\rho u^\sigma\nabla_\nu  u_\mu\nabla_\sigma u_\rho\right)+\cdots \Big\} \, , \label{lNLSM8}
\end{align}
where the ellipsis now denotes multi-trace terms, terms whose leading contribution in an expansion in powers of fields contains more than four Goldstones, and odd intrinsic parity terms.\footnote{The remaining even-parity single-trace terms at eighth order in derivatives are the terms 45-66 and 119-135 listed in the supplemental material {\tt http://home.thep.lu.se/$\sim$bijnens/chpt/basis.pdf} of~\cite{Bijnens:2018lez}.} In what follows we will not need these terms, since we will be calculating the 5- and 6-point functions only up to $\mathcal{O}(p^6)$.

\section{Higher-order Lagrangian for the special galileon} \label{sg}

We now turn our attention to the higher derivative corrections to the special galileon. Our goal is to find the most general action invariant under the SGal symmetries in four space-time dimensions. These symmetries act on the SGal field as~\cite{Hinterbichler:2015pqa}
\begin{subequations} \label{galileon symmetries}
	\begin{align}
	\delta_c\pi&=c \, \\
	\delta_b\pi&=b_\mu x^\mu \ , \\
	\delta_s\pi&=s_{\mu\nu}x^\mu x^\nu + \alpha^2 s^{\mu\nu}\partial_\mu\pi \partial_\nu\pi \ ,
	\end{align}
	\label{sym}
\end{subequations}
where $c$, $b_\mu$, and $s_{\mu\nu}$ (the latter being traceless and symmetric) are the parameters of the symmetry transformations, while $\alpha$ is a constant that is convenient to introduce for normalization purposes. If $\pi$ is a canonically-normalized field, then $\alpha$ must have dimensions of $\text{(mass)}^{-3}$, {\it i.e.} $\alpha \equiv 1/\Lambda^3$. While ordinary galileons are only invariant under the first two shift symmetries~\cite{Nicolis:2008in}, the special galileon also satisfies the third one~\cite{Hinterbichler:2015pqa}. The fact that $\delta_s\pi \sim x^2$ endows the leading order special Galileon field with a particularly soft infrared behavior~\cite{Cheung:2014dqa}.

\subsection{Coset construction and lowest-order effective Lagrangian}

As is the case for any theory with non-linearly realized symmetries, the SGal theory can also be obtained from a coset construction. This was first carried out in four dimensions in~\cite{Bogers:2018zeg}, and later extended to arbitrary dimensions in \cite{Garcia-Saenz:2019yok}. We will now briefly review this construction, and in the next subsection we will use it to systematically write down higher-derivative corrections in four dimensions.

The symmetry transformations \eqref{galileon symmetries} are associated, respectively, with some generators $C, Q_a$, and $S_{ab}$, which, together with the generators of the Poincar\'e group ($P_a$ and $J_{ab}$) satisfy the following algebra~\cite{Hinterbichler:2015pqa}:
\begin{subequations} \label{special Galileon algebra}
	\begin{align}
	&[P_a,Q_b]=\eta_{ab}C \ , \\
	&[J_{ab},Q_\nu]=\eta_{ac}Q_b-\eta_{bc}Q_a \ ,  \\
	&[P_a,S_{bc}]=\eta_{ab}Q_c+\eta_{ac}Q_b-\frac{1}{2}\eta_{bc}Q_a \ , \\
	&[Q_a,S_{bc}]= - \alpha^2\left(\eta_{ab}P_c +\eta_{ac}P_b-\frac{1}{2}\eta_{bc}P_a\right) \ ,  \\
	&[S_{ab},S_{cd}]= \alpha^2 \left(\eta_{ac}J_{bd}+\eta_{bc}J_{ad}+\eta_{bd}J_{ac}+\eta_{ad}J_{bc}\right) \ ,  \\ 
	&[J_{ab},S_{cd}]=\eta_{ac}S_{bd}-\eta_{bc}S_{ad}+\eta_{ad}S_{cb}-\eta_{bd}S_{ca} \ .
	\end{align}
\end{subequations}
The coset parametrization is, as usual, the most general symmetry transformation that is realized non-linearly. In this case, this reads
\begin{equation}
\Omega=e^{x^a P_a}  e^{\pi C} e^{\xi^a Q_a} e^{\frac{1}{2}\sigma^{ab}  S_{ab}} \ .
\end{equation}
The generators of Lorentz transformations, $J_{ab}$, are instead realized linearly, which means that Lorentz invariance of the Lagrangian will be manifest. The Maurer-Cartan form can be calculated using the algebra \eqref{special Galileon algebra}. It takes the form 
\begin{equation}
\Omega^{-1} \ud \Omega = \omega^a_P P_a + \omega^a_Q Q_a + \omega_C C + \frac{1}{2} \omega^{ab}_J J_{ab} + \frac{1}{2} \omega^{ab}_S S_{ab} \ ,
\end{equation}
with 
\begin{align}
\omega_P^a&=E^a{}_\mu \ud x^\mu=(\cos\alpha\sigma)^{ab} \ud x_b - \,\alpha\,(\sin\alpha\sigma)^{ab} \ud \xi_b\\
\omega_C&=\nabla_a\pi \, E^a{}_\mu  \ud x^\mu=\ud \pi+\xi_a\ud x^a \\
\omega_Q^a&=\nabla_c\xi^{a} E^c{}_\mu\ud x^\mu=\frac{1}{\alpha}(\sin \alpha\sigma)^{ab}\ud x_b +(\cos \alpha\sigma)^{ab}\ud \xi_b  \\
\omega_S^{ab}&=\nabla_c\sigma^{ab} E^c{}_\mu \ud x^\mu= \frac{1}{\alpha}\left[\Sigma^{-1} \sin{\alpha\Sigma)}\right]^{ab}_{cd} \ud\sigma^{cd}\\
\omega_J^{ab}&= A_c^{ab} E^c{}_\mu \ud x^\mu=\left[\Sigma^{-1} \left(\cos \alpha\Sigma -1\right)\right]^{ab}_{cd} \ud\sigma^{cd} \ ,
\end{align}
and $\Sigma_{cd}^{ab}\equiv \sigma_{c}^a \delta_d^b  - \sigma_d^b \delta_c^a$. Notice that, despite appearances, these building blocks only depend on even powers of $\alpha$. This is because the algebra depends on $\alpha^2$, not on $\alpha$. Moreover, one can always eliminate $\alpha^2$ from the algebra by an appropriate rescaling of the generators, and therefore only its sign is really physical.

Since we are considering a space-time algebra, the number of broken symmetries does not correspond to the number of Goldstone bosons, and we can apply inverse Higgs constraints that allow us to eliminate some of these modes. In particular, we can demand that
\begin{subequations}
	\begin{eqnarray}
	&\nabla_a \pi =0 \ , & \\
	& \nabla_a\xi_{b} +\nabla_b\xi_{a}-\frac{1}{2}\eta_{ab}  \nabla_c\xi^{c}=0 \  , & \label{SG IH 2}
	\end{eqnarray}
\end{subequations}
and solve these equations to express $\xi_a$ and $\sigma_{ab}$ in terms of derivatives of $\pi$ as follows~\cite{Garcia-Saenz:2019yok}:
\begin{subequations}
	\begin{align}
	\xi_a &=-\partial_a\pi  , \\
	\sigma_{ab} &= \frac{1}{\alpha} \left[(\tan^{-1}\alpha \,\partial\partial\pi)_{ab} -\frac{\eta_{ab}}{4} \, (\tan^{-1}\alpha\,\partial\partial\pi)^c{}_c \right] \ .
	\end{align}
\end{subequations}
Of course, this simply reflects the fact that we only need a single field $\pi$ to non-linearly realize the special Galileon symmetries, as shown in Eq. \eqref{galileon symmetries}.

At lowest order in the derivative expansion, the Lagrangian for any Galileon field (not just the special one) is invariant under the symmetries only up to a total derivative, {\it i.e.} the leading terms are WZW terms~\cite{Goon:2012dy}. In the particular case of the special Galileon, other than the tadpole, there is only one such term. Following the standard procedure to write down WZW terms~\cite{DHoker:1994ti}, it can be built by considering the exact 5-form~\cite{Bogers:2018zeg,Garcia-Saenz:2019yok}
\begin{equation}
d \beta \equiv \sum_{n \text{ even}} \frac{1}{4} \, \omega_C  \wedge \left( \frac{1}{6} \omega_Q^{a} \wedge \omega_Q^{b} \wedge \omega_Q^{c}\wedge \omega_Q^{d} + \omega_Q^{a} \wedge \omega_Q^{b} \wedge \omega_P^{c}\wedge \omega_P^{d} + \frac{1}{6} \omega_P^{a} \wedge \omega_P^{b} \wedge \omega_P^{c}\wedge \omega_P^{d} \right) \epsilon_{abcd} . 
\end{equation}
Up to an overall constant, the coefficient of the 4-form $\beta$ is equal to the leading order Lagrangian for the special galileon~\cite{Hinterbichler:2015pqa}:
\begin{equation}
\lag_\text{SGal}=\frac{1}{2}\left(\partial\pi\right)^2-\frac{\alpha^2}{12}\left(\partial\pi\right)^2\left[\left(\square\pi\right)^2-\left(\partial_\mu\partial_\nu \pi\right)^2\right] \ . \label{sgL}
\end{equation}

\subsection{Higher-derivative corrections}

Higher order terms in the Lagrangian for the special Galileon are exactly invariant under all the symmetries. These can be built using the following ingredients:
\begin{enumerate}
	\item The components of the Goldstones' covariant derivatives that have not been set to zero by imposing inverse Higgs constraints. A priori, these would be $\nabla_a \xi^a, \nabla_{[a} \xi_{b]}$, and $\nabla_a \sigma^{bc}$. However, after solving the inverse Higgs constraint \eqref{SG IH 2} one finds that $\nabla_{[a} \xi_{b]} = 0$~\cite{Garcia-Saenz:2019yok}. Thus, the only non-trivial components are $\nabla_a \xi^a$ and $\nabla_a \sigma^{bc}$.
	\item Additional covariant derivatives, which according to Eq. \eqref{coset cov dev} are defined using the unbroken Lorentz generators\footnote{With normalization conventions for the generators, $(J_{ab})_{cd} = \eta_{ac} \eta_{bd}-\eta_{ad}\eta_{bc}$.}  as $\nabla_a\equiv(E^{-1})_a{}^\mu\partial_\mu+ \frac{1}{2} A_a^{bc}J_{bc}$.
	\item The determinant of the coset vierbein $E_\mu{}^a$, to make the integration measure in the action invariant under the non-linearly realized symmetries.
\end{enumerate}
Based on the building blocks listed above, we conclude that the most general action for the special Galileon must take the form
\begin{equation}
S = \int d^4 x \, \Big[ \mathcal{L}_{SGal} + \det (E) \, \Delta \mathcal{L} ( \nabla_a \xi^a, \nabla_a \sigma^{bc}, \nabla_a) \Big] \ ,
\end{equation}
where $\Delta \mathcal{L}$ contains all possible Lorentz-invariant combinations of its arguments. 

In Sec. \ref{sec: SG amplitudes}, we will use this Lagrangian to study the scattering amplitudes of  the special Galileon. In order to be exactly invariant under the standard galileon symmetry, all higher derivative corrections in $\Delta \mathcal{L}$ must have at least two derivatives acting on each field $\pi$. Hence, we will write $\Delta \mathcal{L} = \sum_{n = 0}^\infty \Delta \mathcal{L}^{(2n)}$, where the superscript $2n$ refers to the number of additional derivatives. For example, keeping in mind that $\nabla \xi \sim \mathcal O (0)$ and $\nabla \sigma \sim \mathcal O (1)$ according to this derivative counting, the first two contributions to $\Delta \mathcal{L}$ are 
\begin{align}
\Delta \mathcal{L}^{(0)} &= A(\nabla \xi) \ , \\
\Delta \mathcal{L}^{(2)} &=  B_1 (\nabla \xi) \nabla_b \nabla^b\nabla_a \xi^a + B_2 (\nabla \xi) \nabla_b \nabla_a \xi^a \nabla^b \nabla_c \xi^c   \label{delta L 2 galileon} +B_3 (\nabla \xi) \nabla^a\nabla^b \sigma_{ab}\\
& \qquad\quad  + [ B_4 (\nabla \xi) \eta^{ab} \eta^{de}\eta^{cf}+B_5 (\nabla \xi) \eta^{ad} \eta^{be}\eta^{cf} +B_6 (\nabla \xi)\epsilon^{abde} \eta^{cf}] \nabla_a \sigma_{bc} \nabla_d \sigma_{ef} \nonumber 
\end{align}
where $A$ and $B_i$ are functions of $\nabla_a \xi^a$ that admit a Taylor expansion around zero. Notice that higher coset covariant derivatives cannot be integrated by parts as one might naively expect. Therefore, say, the first two terms in \eqref{delta L 2 galileon} are independent structures that are both allowed by the symmetries. 

In order to calculate the 4-point function at $\mathcal{O}(p^{12})$ we only need to consider operators in $\Delta \mathcal{L}^{(0)}$, $\Delta \mathcal{L}^{(2)}$ and $\Delta \mathcal{L}^{(4)}$ that can give rise to quartic self-interactions. To calculate the 5-point function at $\mathcal{O}(p^{10})$ and the 6-pt function at $\mathcal{O}(p^{12})$, we also include in $\Delta \mathcal{L}^{(0)}$ those operators that contribute at fifth and sixth order in the fields. To obtain explicit expressions for the interaction vertices we will use  the following expansions in powers of~$\pi$:
\begin{align}
\det(E)=& \,\, 1+\frac{\alpha^2}{2}\left\{\left[(\partial\partial\pi)^2\right]-\frac{1}{4}(\square\pi)^2\right\}+\frac{\alpha^4}{4!}\bigg\{3\left[(\partial\partial\pi)^2\right]^2-6\left[(\partial\partial\pi)^4\right] \\ 
&\quad + \frac{5}{32}(\square\pi)^4-\frac{3}{2}\left[(\partial\partial\pi)^2\right](\square\pi)^2+2\left[(\partial\partial\pi)^3\right](\square\pi)\bigg\} +\frac{\alpha^6}{6!} \bigg\{ -\frac{17}{128} (\square \pi)^6  \nonumber \\
&\quad  +120 [ (\partial \partial \pi)^6 ] -36 \square \pi [ (\partial \partial \pi)^5 ] +\frac{45}{2}  (\square \pi)^2 [ (\partial \partial \pi)^4 ] -\frac{25}{4}  (\square \pi)^3 [ (\partial \partial \pi)^3 ]\nonumber  \\
&\quad  - 10 [ (\partial \partial \pi)^3 ]^2 + \frac{75}{32} (\square \pi)^4 [ (\partial \partial \pi)^2 ] - 90 [ (\partial \partial \pi)^2 ] [ (\partial \partial \pi)^4 ]\nonumber \\
&\quad + 30 \square \pi [ (\partial \partial \pi)^2 ] [ (\partial \partial \pi)^3 ] - \frac{45}{4} (\square \pi)^2 [ (\partial \partial \pi)^2 ]^2 +15 [ (\partial \partial \pi)^2 ]^3 \bigg\} + \mathcal{O}(\pi^8) \ , \nonumber \\
\nabla_a\xi^{a}=
& - \square \pi +\frac{1}{3} \, \alpha^2 \left(\left[(\partial\partial\pi)^3\right]-\frac{1}{16}(\square \pi)^3\right)-\frac{1}{5} \, \alpha^4 \Big(\left[(\partial\partial\pi)^5\right]-\frac{5}{48}(\square \pi)^2\left[(\partial\partial\pi)^3\right] \nonumber  \\
&+\frac{1}{384}(\square \pi)^5\Big)+\mathcal{O}(\pi^7) \ ,  \\
\nabla_c\sigma_{ab}=
& \,\partial^{c}\partial^{b}\partial^{a}\phi -  \frac{1}{4} \eta^{ab} 
\partial^{c} \square \phi 
+ \alpha^2 \bigg\{- \frac{1}{2} 
\partial_{d}\partial^{c}\phi \partial_{h}\partial^{d}\phi 
\partial^{h}\partial^{b}\partial^{a}\phi -  \frac{1}{2} \partial_{d}
\partial^{b}\phi \partial_{h}\partial^{d}\phi 
\partial^{h}\partial^{c}\partial^{a}\phi \nonumber \\ 
& \, -  \frac{1}{2} \partial_{d}
\partial^{a}\phi \partial_{h}\partial^{d}\phi 
\partial^{h}\partial^{c}\partial^{b}\phi + \frac{1}{4} \eta^{ab}
\partial_{d}\partial^{f}\phi \partial_{h}\partial_{f}\phi 
\partial^{h}\partial^{d}\partial^{c}\phi+ \frac{1}{32} \partial^{c}
\partial^{b}\partial^{a}\phi (\square \phi)^2
\nonumber \\ 
& \, -  \frac{1}{128} \eta^{ab} (\square\phi)^2 
\partial^{c}\square\phi  + \frac{1}{8} \eta^{ab} 
\partial_{d}\partial^{e}\phi \partial_{e}\partial^{c}\phi 
\partial^{d}\square\phi \bigg\} + \mathcal{O}(\pi^5) ,
\end{align}
where $[ \cdots ]$ denotes a trace over the Lorentz indices. For example, $\left[(\partial\partial\pi)^3\right]$ stands for $\partial^\mu\partial_\nu\pi\partial^\nu\partial_\rho\pi\partial^\rho\partial_\mu\pi$.  The number of operators in $\Delta \mathcal L^{(2n)}$ grows quickly with $n$. However, this state of affairs simplifies considerably when one realizes that the only non-vanishing tree-level contributions to $n$-point on-shell amplitudes can come from $\mathcal{O}(\pi^m)$ vertices with $m \leqslant n$ and with at most $n-m$ powers of $\square \pi$. This is because factors of $\square \pi$ vanish on-shell, and therefore can be ignored unless they are acting on internal lines of Feynman diagrams. As a result, the only operators that are relevant to our calculations are 
\begin{align}
\Delta \mathcal{L}^{(0)} &\to a_0 + a_1 \nabla_a \xi^a + \frac{a_2}{2} (\nabla_a \xi^a)^2 \\
\Delta \mathcal{L}^{(2)} & \to b_1 \nabla_a \sigma_{bc} \nabla^a \sigma^{bc} + b_2 \nabla_a \sigma_{bc} \nabla^c \sigma^{ab} \\
\Delta \mathcal{L}^{(4)} & \to c_{1}^{} \nabla_{a}\sigma^{de} \nabla^{a}\sigma^{bc} \nabla_{b}\sigma_{d}{}^{f} \nabla_{c}\sigma_{ef} + c_{2}^{} \nabla_{a}\sigma_{b}{}^{d} \nabla^{a}\sigma^{bc} \nabla_{c}\sigma^{ef} \nabla_{d}\sigma_{ef} + c_{3}^{} (\nabla_{a}\sigma_{bc} \nabla^{a}\sigma^{bc})^2 \nonumber \\
& \qquad + c_4 \nabla_a \nabla_b \sigma_{cd} \nabla^a \nabla^b  \sigma^{cd}  + c_{5}^{} \nabla_{a}\sigma_{cd} 
\nabla^{a}\nabla^{b}\nabla_{b}\sigma^{cd} + c_{6}^{} 
\nabla_{a}\sigma_{cd} \nabla^{a}\nabla^{b}\nabla^{c}\sigma_{b}{}^{d} \nonumber \\
& \qquad + c_{7}^{} \nabla_{a}\sigma_{bd} 
\nabla^{a}\nabla^{b}\nabla^{c}\sigma_{c}{}^{d} + c_{8}^{} 
\nabla^{a}\nabla_{a}\nabla^{b}\sigma^{cd} \nabla_{b}\sigma_{cd} + 
c_{9}^{} \nabla^{a}\nabla^{b}\nabla_{a}\sigma^{cd} 
\nabla_{b}\sigma_{cd} \nonumber \\
& \qquad + c_{10}^{} 
\nabla^{a}\nabla^{b}\nabla^{c}\sigma_{a}{}^{d} \nabla_{b}\sigma_{cd} \ .
\end{align}

A few comments are in order at this point. First, we have omitted from $\Delta \mathcal{L}^{(4)}$ those operators that, despite being linearly independent from the ones shown, would yield redundant interactions at quartic order. Second, it is easy to see that this Lagrangian will give rise to higher derivative corrections to the 2-point function of the form $\pi \, \square^n \pi$. From an EFT viewpoint, these terms should be treated perturbatively as one does with any other higher-derivative interaction, and not used to modify the propagator. (See for instance footnote 1 in~\cite{Weinberg:2008hq} for a brief discussion of this point.) Finally, the second operator in $\Delta \mathcal{L}^{(0)}$ gives rise to a cubic vertex. Nevertheless, this vertex does not contribute to the scattering amplitudes since it vanishes when one leg is on-shell\footnote{We thank Jiri Novotny and Filip Preucil for making us aware of this fact.}. Similarly, higher derivative 3-point vertices that do not vanish when one leg is on-shell (such as the 8th derivative ones arising from $\nabla_b\nabla^b\nabla_a \xi^a$, and $\nabla^a\nabla^b \sigma_{ab}$), do not spoil the single soft limit due to the large number of momentum factors involved in them.  In fact, it has been argued in \cite{Preucil:2019nxt} that using the leading order equations of motion one can show that these operators shouldn't contribute to the scattering amplitudes.

\section{Compatibility with the double copy} \label{sec: double copy}
In this section we will analyze the corrections to the 4-, 5- and 6-point amplitudes of the NLSM and SGal that follow from the higher derivative operators introduced in the previous two sections. We will be particularly interested in understanding the extent to which these corrections are compatible with the double copy procedure.

\subsection{NLSM Scattering Amplitudes} \label{Sec: NLSM double copy}

We do this first for the NLSM introduced in Sec. \ref{sec: higher der NLSM} are compatible with the double copy procedure. To this end, we will expand the operators in Eqs. \eqref{lNLSM4}, \eqref{NLSM WZW}, \eqref{lNLSM6}, and \eqref{lNLSM8} in powers of fields, and use the resulting interactions to calculate the 4-, 5-, and 6-point on-shell amplitudes for the Goldstones. 

An important point to notice is that, in order to be compatible with color-kinematics duality, the color structure of the scattering amplitudes must satisfy Jacobi identities. This is a necessary but not sufficient condition to guarantee the existence of the double copy, since one also needs the correct kinematic behavior. Focusing on the color factors arising from the higher-order corrections to the NLSM, one sees that multi-trace color factors can arise at tree level. Crucially, for a general $SU(N)$ group these are not related to the single-trace color factors, and the color factors associated with multi-trace operators in the Lagrangian would not necessarily satisfy Jacobi identities. Whether or not these terms are compatible with a (modified) double copy procedure is still unknown. For examples in which multi-trace terms are analyzed and generalized BCJ relations are considered see \cite{Nandan:2016pya,Schlotterer:2016cxa,Du:2017gnh}. From now on, we will neglect the multi-trace terms, noting that, as we discussed in Sec. \ref{sec: higher der NLSM}, the large-$N$ limit makes our approach self-consistent.  Restricting our attention to single-trace operators, we see that the corresponding amplitudes can be cast in the form
\begin{equation} \label{color ordered def}
\mathcal{A}_n^{\alpha_1, \dots, \alpha_n} (p_1, \dots , p_n) = \sum_{\sigma \in S_{n-1} / Z_{n-1}} \text{Tr} [X^{\alpha_1} X^{\alpha_{\sigma(2)}} \dots X^{\alpha_{\sigma(n)}}] A_n (p_1, p_{\sigma(2)}, \dots, p_{\sigma(n)}),
\end{equation}
where $S_{n-1}$ is the set of all possible permutations of $n-1$ objects, whereas $Z_{n-1}$ is the subset of cyclic permutations. The quantities $A_n (p_1, \dots, p_n)$ are known as color-ordered (or partial) $n$-point amplitudes in the trace basis. In what follows, we denote these quantities simply as $A_n [1, \dots, n]$. Our explicit results for $A_4$ up to $\mathcal{O} (p^8)$, and for $A_5$ and $A_6$ up to $\mathcal{O} (p^6)$ can be found in Appendix \ref{amp}. 
  
The existence of a double copy also requires the color-ordered amplitudes to have a special kinematic structure. In fact, we must demand that they satisfy the KK~\cite{Kleiss:1988ne} and BCJ~\cite{Bern:2008qj} relations, which can be expressed respectively as
\begin{align}
A_n[1,\{\alpha\},n,\{\beta\}]=(-1)^{|\beta|}\sum_{\sigma\in\text{OP}(\{\alpha\},\{\beta^T\})}A_n[1,\sigma,n] \ , \\
\sum_{i=3}^{n} \, \sum_{j=3}^{i}s_{2j} A_n[1,3,\dots,i,2,i+1,\dots,n]=0 \ .
\end{align}
In these relations, $\{\alpha\}$ and $\{\beta\}$ are subsets of the external particle labels, $|\beta|$ is  the number of  elements in the subset $\{\beta\}$, the superscript $T$ denotes the reverse ordering, OP denotes ordered permutations\footnote{These are permutations that preserve the ordering of the set $\{\alpha\}\cup \{\beta^T\}$---see {\it e.g.} page 35 of~\cite{Elvang:2015rqa} for some useful examples.},  and $s_{2j}=(p_2\cdot p_j)^2$ is the usual Mandelstam variable. The fact that the leading order NLSM amplitudes arising from the Lagrangian in Eq. \eqref{lNLSM} satisfy these relations was shown in \cite{Chen:2013fya}.  Imposing that the conditions above are satisfied places constraints on the dimensionless coefficients that appear in Eqs. \eqref{lNLSM4}, \eqref{NLSM WZW}, \eqref{lNLSM6} and \eqref{lNLSM8}, as we will now discuss.  

Let us start by considering the color-ordered 4-point amplitude. The most general form it can take while  satisfying the KK and BCJ relations up to eighth order in derivatives is~\cite{Elvang:2018dco}
\begin{equation} \label{NLSM A4}
A_4[1,2,3,4]=\frac{C_2}{F^2}\, t+\frac{C_6}{F^6}\, t(s^2+t^2+u^2)+\frac{C_8}{F^8}\, t(stu)+\cdots \ ,
\end{equation}
where $s, t$ and $u$ are the usual Mandelstam variables, and $C_i$ are constants with the subscript ``$i$'' labeling the powers of momenta in the corresponding term.  As we already alluded to in the introduction, this amplitude corresponds to that of the Abelian Z-theory~\cite{Carrasco:2016ldy}. The first term in particular follows directly from the lowest order NLSM Lagrangian in Eq.~\eqref{lNLSM}. 

We would like to understand what constraints need to be imposed on the coefficients of higher order corrections to recover an amplitude of the form~\eqref{NLSM A4}. At the 4-derivative level, the contribution arising from the terms in Eq. \eqref{lNLSM4} is of the form
\begin{equation}
	A_4 \propto \frac{c_1}{F^4} t^2  + \frac{c_2}{F^4}  \left(s^2+ st +\frac{t^2}{2}\right).
\end{equation}
This satisfies the KK relations above if $c_1=-c_2$, but the BCJ relation cannot be satisfied. We must therefore set $c_1=c_2=0$, which is consistent with the fact that \eqref{NLSM A4} does not contain any term quartic in momenta. Although there is no $1/F^4$ correction that is compatible with color-kinematics duality, it is interesting to point out that there exists a $1/F^4$ correction that satisfies the NLSM double soft limit and reads $A_4  \propto st/F^4$~\cite{Rodina:2018pcb}. One should note that  this amplitude cannot be obtained from Eq.\eqref{lNLSM4}.  When it comes to the 6- and 8-derivative corrections, one can show that they satisfy both the KK and BCJ relations only if 
\begin{equation*}
d_3=2(d_1+d_2),  \ \ \  g_1+g_2=0 \, \ \  \text{ and } \ \  g_3+2g_4=0 \ .
\end{equation*}

Moving on to the 5-point amplitude, we must require that all the contributions with less than 14 derivatives vanish. This is because, as discussed in the introduction, the leading color-ordered 5-point amplitude that is compatible with color-kinematics duality is known to have 14 derivatives~\cite{Elvang:2018dco}. This means that the coefficient in front of the Wess-Zumino term must vanish. Similarly, we must have $e_1 = e_2=0$ in Eq.~\eqref{lNLSM6}. 

It is also interesting to explore whether the 14th derivative order 5-point amplitude which is compatible with color-kinematics duality can be obtained from a Lagrangian satisfying the NLSM symmetries. In order to make some progress towards this question, we will make a few extra assumptions. Assuming that the pions are pseudoscalars and that  the theory is invariant under Parity, $\phi^a(t,\mathbf{x})\rightarrow-\phi^a(t,-\mathbf{x})$; it has been shown that only terms with odd number of Levi-Civita tensors have odd numbers of Goldstones \cite{Witten:1983tw}. In this case, the general  form of the 5-point NLSM amplitude is 
\begin{equation} \label{NLSM A5}
A_5[1,2,3,4,5]=\epsilon_{\mu\nu\lambda\rho}\ p_1^\mu p_2^\nu p_3^\lambda p_4^\rho \, \Gamma (p_1,p_2,p_3,p_4,p_5) \ ,
\end{equation}
where $\Gamma$ is some scalar function constructed from the Goldstone momenta. Rather than calculating this amplitude explicitly by considering all possible terms in the Lagrangian that could contribute up to $\mathcal{O} (p^{14})$, we have followed a different approach. We have constructed the most general 5-point amplitude of the form \eqref{NLSM A5} at $\mathcal{O} (p^{14})$ and verified explicitly that it cannot satisfy the  KK and  BCJ relations. This means that the term found in~\cite{Elvang:2018dco} cannot be invariant under the NLSM symmetries when considering pseudoscalar pions in a parity invariant theory. In a more general NLSM setting, a 5-point amplitude at 14th derivative order could arise from two different kinds of operators. The contact terms could come from operators of the form $\nabla^4(\nabla u)^5$ and $\nabla^8(\nabla u)^3$. Meanwhile, the pole terms would come from a 4-point vertex of order $p^n$ and a 3-point vertex of order $p^{16-n}$ which comes from an operator $\nabla^{10-n}(\nabla u)^3$. The calculation of the amplitude arising from these terms seems intractable and would not be perform here.

Finally, we consider the 6-point amplitude up to $\mathcal{O}(p^6)$.  The 4-derivative contribution to this amplitude vanishes by virtue of the requirements  already  imposed  on  the  4-point amplitude. Meanwhile, the 6-derivative contribution arising from Eq.~\eqref{lNLSM6} satisfies the  KK and BCJ relations if
\begin{equation*}
d_3=2(d_1+d_2),  \ \ \   f_3 = -8 d_1,  \  \ \  f_4 =-\frac{8}{3} (2 d_1 + 5 d_2), \ \  \text{and} \ \   f_5 =
8 (d_1 + 2 d_2) \ .
\end{equation*} 
Furthermore, this amplitude is equal to the Abelian Z-theory result if
\begin{equation*}
d_2 = -\frac{1}{64} - \frac{d_1}{4}, \ \ \   f_1 = -\frac{1}{12} + 4 d_1, \ \ \text{and}  \ \ f_2 = \frac{1}{8} + 2 d_1 \ .
\end{equation*}

\subsection{SGal Scattering Amplitudes} \label{sec: SG amplitudes}

We now consider the scattering amplitudes arising from the higher derivative special galileon Lagrangian. Explicit expression for the  4- and 6-point amplitudes up to $\mathcal{O}(p^{12})$  can be found in Appendix \ref{SGA}. 

Before turning our attention to the double copy, it is worth discussing briefly the single soft limit of the 4-point amplitude, which reads
\begin{align}
\mathcal{A}_4&=-\frac{2}{\Lambda^6}  stu+ \frac{1}{\Lambda^{10}} \left(\frac{ b_1+b_2}{40}\right)\left(s^5+t^5+u^5\right)+\frac{(c_1-6c_2)}{2^6 \Lambda^{12}}s^2t^2u^2 \nonumber \\
& \qquad \qquad +\frac{1}{\Lambda^{12}} \left(c_2+c_3+c_4\right)\left(s^6+t^6+u^6\right)+\mathcal{O}(p^{14}) \ .
\end{align}
The fact that the term with 8 derivatives is not present comes from a non-trivial cancellation happening in $\det(E)$. This cancellation is crucial to have a soft theorem with degree $\sigma=3$. By comparing our results with the ones obtained with the soft bootstrap method~\cite{Elvang:2018dco}, we find agreement. The term $s^6+t^6+u^6$  receives contributions proportional to the coefficients $c_2$, $c_3$  and  $c_4$. We have checked explicitly that these coefficients can enter the 6-point amplitude without affecting the enhanced soft limit. In fact, the authors of \cite{Elvang:2018dco} agree that such term is possible.\footnote{Private communication with the authors of  \cite{Elvang:2018dco}.} 

We note that the leading contribution to the 5-point amplitude arising in the soft bootstrap case at $\mathcal{O}(p^{14})$ does  not come from a Lagrangian with special galileon symmetry. This amplitude could arise from terms such as $\epsilon_{bcde}\nabla^b \nabla^c \nabla^d \nabla^e  \nabla^a\xi_a$ and $\epsilon_{bcde}\nabla^b \nabla^c \nabla^d \nabla^a  \sigma^{e}_{a}$, nevertheless the resulting amplitude vanishes. As a matter of fact, up to the 14th derivative order we have checked that all contributions to the 5-point amplitude vanish. This is consistent with the results found in \cite{Preucil:2019nxt}. While a proof for all derivative orders is unavailable, these results seem to indicate that odd-point amplitudes arising from a special galileon invariant theory vanish on-shell.

We now compare the special galileon amplitudes with the double copy of the most general colored ordered scalar amplitudes satisfying the KK and BCJ relations. While the even-point amplitudes correspond to the NLSM ones with dimensionless coefficients constrained as in the previous section, we will also include for completeness a 5-point amplitude $A_5^*$ at 14th derivative order which does not arise from a parity invariant NLSM Lagrangian, and yet enjoys the same single soft limit. Using these building blocks, we can construct the KLT double copy as follows:
\begin{align}
A^\text{DC}_{4}(1,2,3,4)&=-   s_{12} \ A_{4}[1,2,3,4] \  A_{4}[1,2,4,3] \ ,\\
A^\text{DC}_{5}(1,2,3,4,5)&= s_{12} s_{34} A^*_{5}[1,2,3,4,5] \  A^*_{5}[2,1,4,3,5]+\mathcal{P}(2,3)  \ , \\
A^\text{DC}_{6}(1,2,3,4,5,6)&=-s_{12} s_{45} A_{6}[1,2,3,4,5,6]\left(s_{35} A_{6}[1,5,3,4,6,2]\right. \nonumber \\
& \qquad \qquad \qquad +\left(s_{34}+s_{35}\right) A_{6}[1,5,4,3,6,2] )+\mathcal{P}(2,3,4) \ ,
\end{align}
where $\mathcal{P}(2,3)$ denotes all the permutations of the momenta $p_2$ and $p_3$, and so on. 

By comparing the scattering amplitudes obtained from the SGal Lagrangian with the ones obtained from the KLT double copy shown above, we find that we need to set $c_2=c_3=c_4=0$ since the term $s^6+t^6+u^6$ does not arise in the double copy. This shows that, by constraining the coefficients of the allowed operators in both the NLSM and the SGal, we can maintain their relation through the double copy. At this point, we lack a compelling argument which explains these constraints, but we discuss some possibilities in the next section. To conclude, we should mention that the leading order 5-point amplitude that can be obtained as the double copy of a colored scalar arises at  $\mathcal{O}(p^{32})$.  Understanding whether this could arise from a special galileon invariant action is beyond the scope of this paper, but it would seem implausible given that all the computed odd-point amplitudes vanish on-shell. 

\section{Discussion and conclusions}  \label{conc}

We have constructed the higher derivative Lagrangians for both the non-linear sigma model and the special galileon by using building blocks given by the coset construction. The explicit form of these Lagrangians would be particularly important to calculate the radiation emitted at higher orders in the context of the classical perturbative double copy of~\cite{Carrillo-Gonzalez:2018pjk}. Here, however, we focused on the on-shell scattering amplitudes arising from these shift-symmetric Lagrangians, and discussed their compatibility with the double copy. Without the Lagrangian realization, we would not be able to tell if a scattering amplitude arises from a shift symmetric theory.

For the NLSM, we have analyzed whether it is possible to obtain amplitudes which satisfy the KK and BCJ relations by imposing constraints on the dimensionless coefficients appearing in our Lagrangian. We showed that this is possible for the 4-point amplitude up to the 8th derivative order. On the other hand, we found that the leading order 5-point amplitude which satisfies KK and BCJ relations does not arise from a theory invariant under the $U(N)$-NLSM symmetries and parity. Nevertheless, it is still possible that a Lagrangian which does not involve any Levi-Civita tensors could give rise to such 5-point amplitude. For the 6-point amplitude, we have found that the NLSM amplitudes up to $\mathcal{O}(p^6)$ can satisfy these relations provided the coupling constants satisfy certain constraints.  At this stage, we are not aware of any symmetry that would enforce these constraints. Moreover, we have not explored whether these tunings happen to be technically natural. However, it is worth noticing that the constrained amplitudes still admit more than one free parameter. In principle, it might seem surprising that an amplitude with more than one free parameter satisfies  the KK and BCJ relations, but we believe that this is due to the fact that, when this happens, the $\sigma=1$ soft limit is trivially satisfied. It is also relevant to mention that the Abelian Z-theory amplitudes correspond to a subset of the constrained NLSM amplitudes involving only one free parameter. When combined, these results show that, at least up to the derivative order we have considered, the most general colored-scalar theory compatible with color-kinematics duality is not merely a subset of the $U(N)$-NLSM.

We have also explicitly constructed the higher-order Lagrangian invariant under the special galileon symmetries, and have used this to understand the disagreement between the different definitions of the special galileon.  It was previously shown that the even-point amplitudes of a scalar field with soft degree $\sigma=3$, except for the $s^6+t^6+u^6$ term, match the scattering amplitudes obtained as the double copy of the most general colored scalar satisfying the KK and BCJ relations \cite{Elvang:2018dco}. In \cite{Elvang:2018dco}, it was also shown that there  is a 5-point amplitude with soft degree $\sigma=3$ but too few momenta to arise from the double copy. This is the first instance in which the definitions of the special galileon based on its single soft limit or the double copy procedure have turned out to be inequivalent; in other words, the most general scalar field amplitudes with a soft degree $\sigma=3$ do not correspond to the double copy of the most general colored scalar satisfying the KK and BCJ relations. In order to restore the equivalence, one could only consider  even-point amplitudes and remove the $s^6+t^6+u^6$ term from the 4-point amplitude. 

In our construction, we are able to constrain the dimensionless coefficients on both the NLSM and the SGal side in order to maintain their relation through the double copy. This is possible since only the even terms (up to the computed derivative orders) on the NLSM side satisfy the KK and BCJ relations, this matches the fact that the only non-vanishing amplitudes of the SGal are the even ones. It would be interesting to analyze the origin on the constraints set on the Wilsonian coefficients of these EFTs. A possibility worth exploring is if these constraints are related to the positivity bounds of EFTs that allow for a local, analytic, unitary UV completion \cite{Adams:2006sv,deRham:2017avq}, or other unitary conditions such as those in \cite{Chen:2019qvr,Green:2019tpt,Huang:2020nqy}.

We have also discussed whether the 5-point amplitude arising as the double copy of the 14th derivative color-ordered amplitude which satisfies  KK and BCJ relations could come from a theory with the SGal symmetries. We do not construct this amplitude since its calculation through Feynman rules seems intractable. Developing amplitude methods along the lines of the soft bootstrap method applied in \cite{Low:2019ynd} that can compute higher order corrections appears to be a more promising approach. Nevertheless, it seems unlikely that odd-point amplitudes arise from the SGal invariant action; a complete proof could follow the lines of the analysis in \cite{Preucil:2019nxt}.
 
As summarized  in Fig.\ref{fig:summary}, the results for both the NLSM and the  SGal higher order amplitudes tell us that the definitions of the exceptional scalar theories based on their symmetries, single soft  limits, or double copy relations are not equivalent beyond leading order. As we  mentioned in the introduction, there are various methods for computing the higher derivative on-shell scattering amplitudes, but only a few that also obtain the corresponding Lagrangians. Given this, it would  be interesting to explore whether the most general higher derivative corrections compatible with the double copy can be obtained as a dimensional reduction of higher-order operators of Yang-Mills theories and gravity, in the spirit of \cite{Cheung:2017yef,Cheung:2017ems,Mizera:2018jbh}.

\acknowledgments
We thank Henriette Elvang, Sebastian Garcia-Saenz, Marios Hadjiantonis, Kurt Hinterbichler, Callum Jones, Austin Joyce, Shruti Paranjape, and David Stefanyszyn for useful discussions. We also thank Jiri Novotny and Filip Preucil for their helpful insights regarding the special galileon amplitudes and for pointing out a mistake in the first draft of this paper.  RP would also like to thank Aleksandr Zheltukhin for bringing Ref. \cite{Volkov:1973up} to his attention. The work of MCG and MT is supported in part by US Department of Energy (HEP) Award DE-SC0013528. The work of RP is partially supported by the National Science Foundation under Grant No. PHY-1915611.

\appendix
\section{Useful identities for simplifying  the NLSM Lagrangian} \label{App Bianchi Levi Civita}

When considering the alternative coset parametrization for the NLSM of section \ref{AltCos}, we have a non-zero connection given by Eq.\eqref{conn}. The geometric structure of the coset space allows us to define a field strength, $\Gamma_{\mu \nu}$,  corresponding to this connection by
\begin{align}
\left[\nabla_{\mu}, \nabla_{\nu}\right] X=\left[\Gamma_{\mu \nu}, X\right] \ , \qquad \Gamma_{\mu \nu}=\frac{1}{4}\left[u_{\mu}, u_{\nu}\right] \ .  
\end{align}
This field  strength satisfies the Bianchi identity
\begin{equation}
\nabla_{\mu} \Gamma_{\nu \rho}+\nabla_{\nu} \Gamma_{\rho \mu}+\nabla_{\rho} \Gamma_{\mu \nu}=0 \ ,
\end{equation}
which is useful in simplifying the NLSM Lagrangians.

On the other hand, there are identities that specifically help us to simplify the odd intrinsic parity terms. These are Levi-Civita identities which follow from the fact that, in 4d, a completely  antisymmetric tensor with 5 indices is  zero, that is, 
\begin{equation}
g^{\alpha \beta} \epsilon^{\gamma \rho \tau \eta}-g^{\alpha \gamma} \epsilon^{\beta \rho \tau \eta}-g^{\alpha \rho} \epsilon^{\gamma \beta \tau \eta}-g^{\alpha \tau} \epsilon^{\gamma \rho \beta \eta}-g^{\alpha \eta} \epsilon^{\gamma \rho \tau \beta}=0 \ .
\end{equation}
Contracting a tensor $T_{\alpha \beta \gamma \rho \tau \eta}$ in every possible way with the one above leads  to the (independent) identities:
\begin{align}
\left( -\tensor{T}{_\alpha^\alpha_{\gamma \rho \tau \eta}}+\tensor{T}{_{\alpha \gamma}^{\alpha}_{\rho\tau\eta}}-\tensor{T}{_{\alpha \gamma \rho}^{\alpha}_{\tau\eta}}+\tensor{T}{_{\alpha \gamma \rho \tau}^{\alpha}_\eta}-\tensor{T}{_{\alpha \gamma \rho \tau \eta}^{\alpha}} \right) \epsilon^{\gamma \rho \tau \eta}=0  \ , \\
\left(+\tensor{T}{_\alpha^\alpha_{\gamma \rho \tau \eta}}+\tensor{T}{_{ \gamma\alpha}^{\alpha}_{\rho\tau\eta}}+\tensor{T}{_{\gamma \alpha \rho}^{\alpha}_{\tau \eta}}-\tensor{T}{_{\gamma \alpha \rho \tau}^{\alpha}_{\eta}}+\tensor{T}{_{\gamma \alpha \rho \tau \eta}^{\alpha}}\right) \epsilon^{\gamma \rho \tau \eta}=0 \ , \\
\left(-\tensor{T}{_{\alpha \gamma}^{\alpha}_{\rho\tau\eta}}+\tensor{T}{_{\gamma \alpha}^{\alpha}_{\rho \tau \eta}}-\tensor{T}{_{\gamma \rho \alpha}^{\alpha}_{\tau \eta}}+\tensor{T}{_{\gamma \rho \alpha \tau}^{\alpha}_{\eta}}-\tensor{T}{_{\gamma \rho \alpha \tau \eta}^{\alpha}}\right) \epsilon^{\gamma \rho \tau \eta}=0  \ , \\
\left(+\tensor{T}{_{\alpha \gamma \rho}^{\alpha}_{\tau\eta}}-\tensor{T}{_{\gamma \alpha \rho}^{\alpha}_{\tau \eta}}+\tensor{T}{_{\gamma \rho\alpha}^{\alpha}_{\tau\eta}}-\tensor{T}{_{\gamma \rho \tau \alpha}^{\alpha}_{\eta}}+\tensor{T}{_{\gamma \rho \tau \alpha \eta}^{\alpha}}\right) \epsilon^{\gamma \rho \tau \eta}=0  \  , \\
\left(-\tensor{T}{_{\alpha \gamma \rho \tau}^{\alpha}_\eta}+\tensor{T}{_{\gamma \alpha \rho \tau}^{\alpha}_{\eta}}-\tensor{T}{_{\gamma \rho \alpha \tau}^{\alpha}_{\eta}}+\tensor{T}{_{\gamma \rho \tau \alpha}^{\alpha}_{\eta}} -\tensor{T}{_{\gamma \rho \tau \eta \alpha}^{\alpha}}\right) \epsilon^{\gamma \rho \tau \eta}=0 \ .
\end{align}
When the tensor $T_{\alpha \beta \gamma \rho \tau \eta}$ is constructed out  of $u^\mu$ and $\nabla^\mu$, these identities can be used  to simplify the NLSM Lagrangian. 

\section{Higher-order NLSM Amplitudes} \label{amp}

In this appendix, we report the explicit expressions for the color-ordered amplitudes arising from the NLSM single-trace interactions in Eqs. \eqref{lNLSM4}, \eqref{NLSM WZW}, \eqref{lNLSM6} and \eqref{lNLSM8}. As a cross-check of our calculations, we have verified that these amplitudes have the correct infrared behavior by computing the double soft limit of the 6-point amplitude. 

In what follows, we will denote by $A_n^{(j)}[1, \dots, n]$ the $\mathcal{O}(p^j)$ contribution to the $n$-point on-shell color-ordered amplitude. With this notation, the 4-point color-ordered amplitude of the NLSM up to the eighth derivative order is given by a sum of the following terms: 
\begin{align}
A_4^{(2)}[1,2,3,4]&=-\frac{2}{F^2}t \ ,  \\
A_4^{(4)}[1,2,3,4]&=\frac{16}{F^4} \left(c_1 t^2  + c_2 \left(s^2+ st +\frac{t^2}{2}\right)\right) \ , \\
A_4^{(6)}[1,2,3,4]&=\frac{32}{F^6}\left(\!\!-\frac{3  d_1}{2} \left(s^2t+st^2+\frac{t^3}{3}\right)+d_2 \  t^3+d_3\ t\left(s^2+ st +\frac{t^2}{2}\right)\!\right)  \!  , \\
A_4^{(8)}[1,2,3,4]&=\frac{16}{F^8}\left(\frac{g_1+g_2}{4}s^2(s^2+2st+t^2)+\frac{g_3}{4}t^2 \left(s^2+ st +\frac{t^2}{2}\right)+\frac{g_4}{4}t^4\right) \ ,
\end{align}
where $s$, $t$, and $u$ are the usual Mandelstam variables defined as
\begin{eqnarray}
s=(p_1+p_2)^2, \qquad\qquad t=(p_1+p_3)^2, \qquad\qquad u=(p_1+p_4)^2 \ .
\end{eqnarray} 
The 5-point partial amplitude up to sixth derivative order is given by:
\begin{align}
A_5^{(4)}[1,2,3,4,5]&=\frac{5c}{F^5} \, \epsilon_{\mu\nu\lambda\rho} p_1^\mu p_2^\nu p_3^\lambda p_4^\rho \ , \\
A_5^{(6)}[1,2,3,4,5]&=\frac{64 e_1}{F^7}\,\epsilon_{\mu\nu\lambda\rho}  p_1^\mu p_2^\nu p_3^\lambda p_4^\rho\left[p_1\cdot(p_3+2p_4)+p_2\cdot(p_4-p_3)\right] \nonumber \\
&+\frac{2 e_2}{3 F^7} \Bigg(32 s_{24}^3+3 \left(14 s_{25}+9 s_{34}+7 s_{35}\right) s_{24}^2+3 \Big(14 s_{25}^2+2 \left(6 s_{34}+7 s_{35}\right) s_{25} \nonumber \\
&-9 s_{34}^2-7 s_{35}^2-6 s_{34} s_{35}\Big)
s_{24}-32 s_{34}^3-10 s_{35}^3+6 s_{25} s_{34}^2-3 \left(5 s_{25}+14 s_{34}\right) s_{35}^2 \nonumber \\
&+30 s_{25}^2 s_{34}+3 \left(5 s_{25}^2-2 s_{34} s_{25}-14 s_{34}^2\right)
s_{35}+s_{23}^2 \left(6 s_{24}+57 s_{25}-51 s_{34}+6 s_{35}\right) \nonumber \\
&+3 s_{23} \Big(16 s_{24}^2+2 \left(13 s_{25}-5 s_{34}+2 s_{35}\right) s_{24}+3 \big(5 s_{25}^2+2 s_{34} s_{25}-7
s_{34}^2-4 s_{35}^2 \nonumber \\
&+4 \left(s_{25}-2 s_{34}\right) s_{35}\big)\Big)+10 s_{25}^3\Bigg) \  .
\end{align}
Finally, the 6-point partial amplitude up to sixth derivative order is:
\begin{align}
A_6^{(2)}&[1,2,3,4,5,6] = \frac{2}{F^4}\Bigg(\frac{s_{13}
	s_{46}}{s_{123}}+\frac{s_{15} s_{24}}{s_{234}}+\frac{s_{26} s_{35}}{s_{345}}-s_{24}-s_{26}-s_{46}\Bigg)\ , 
\end{align}
\begin{align}
A_6^{(4)}&[1,2,3,4,5,6]= \frac{4}{F^6}\Bigg(\!\!c_1 \Big[ s_{14} s_{26}+s_{13} \left(s_{25}+s_{26}\right)+s_{24} s_{35}+s_{15} 
\left(s_{26}+s_{36}\right)+\left(s_{15}+s_{25}\right) s_{46}\nonumber \\
&-\frac{2}{3} (s_{14} s_{26}+s_{13} \left(s_{24}+s_{25}+2 s_{26}\right)+s_{15} \left(2 s_{26}+s_{36}\right)+\left(2 s_{15}+s_{25}+s_{35}\right)s_{46}) \Big] \nonumber \\
&+c_2 \Big[s_{23} s_{45}+\left(s_{13}+s_{14}+s_{24}\right) s_{56} +s_{12} \left(s_{35}+s_{36}+s_{46}+s_{56}\right)\nonumber \\
&-\tfrac{2}{3} (\left(s_{13}+2 s_{14}+s_{24}+s_{34}\right) s_{56}+s_{12} \left(s_{34}+s_{35}+2 s_{36}+s_{46}+2 s_{56}\right))\Big] \nonumber \\
&+\tfrac{2}{3 s_{123}} \left(s_{45}-2 s_{46}+s_{56}\right) \Big[2 \text{c}_1 s_{13} \left(s_{12}+s_{23}\right)+\text{c}_2 \left(\left(s_{12}+s_{13}\right) s_{23} \right.\left. + s_{12} \left(s_{13}+s_{23}\right)\right)\Big]  \nonumber \\
	& +\text{cyc}(1,2,3,4,5,6) \Bigg) \ , 
\end{align}
\begin{align}
& \!\!\!\!\!\!\!\!\! A_6^{(6)}[1,2,3,4,5,6]=\frac{2}{F^8}\Bigg\{32\bigg(3
f_1 \left( s_{16} s_{23} s_{45}+ s_{12} s_{34} s_{56}\right)+2 f_2 ( s_{12} s_{36} s_{45}+ s_{16} s_{34}
s_{52}  \nonumber \\
&+ s_{23} s_{41} s_{56})+f_3 \left( s_{16} s_{24} s_{35}+ s_{12} s_{46} s_{35}+ s_{13} s_{26}
s_{45}+ s_{15} s_{23} s_{46}+ s_{13} s_{24} s_{56}+ s_{34} s_{51} s_{62}\right) \nonumber \\
&+6 f_4\, s_{14} s_{36} s_{52}+2 f_5 \left( s_{14} s_{26} s_{35}+ s_{15} s_{36} s_{42}+ s_{25} s_{31} s_{46}\right)\bigg)
\nonumber  \\
&+\Bigg[\frac{1}{3} \Big(2 d_1 \big((s_{12}-2 s_{13}+s_{14}) s_{56}^2+(s_{41}-2 s_{42}+s_{43}) s_{56}^2+s_{12} (s_{34}-2 s_{35}+s_{36}) (s_{34}+s_{35}+s_{36})\nonumber \\
&+s_{12} (s_{63}-2 s_{64}+s_{65}) (s_{63}+s_{64}+s_{65})\big)+2 d_2 \big((s_{62}-2
s_{63}+s_{64}) s_{51}^2+s_{13} (s_{24}-2 s_{25}+s_{26}) s_{31}\nonumber \\
&+s_{46} (s_{51}-2 s_{52}+s_{53}) (s_{51}+s_{52}+s_{53})+(s_{13}-2 s_{14}+s_{15})
(s_{13}+s_{14}+s_{15}) s_{62}\big)\nonumber \\
&+d_3 \big((s_{13}+s_{14}+s_{15}) s_{16} (s_{23}-2
s_{24}+s_{25})+(s_{13}-2 s_{14}+s_{15}) s_{16} (s_{23}+s_{24}+s_{25}) \nonumber \\
&+2 s_{13} (s_{14}-2 s_{15}+s_{16}) s_{32}+2 s_{51} (s_{52}-2 s_{53}+s_{54})
s_{61}+s_{45} (s_{51}+s_{52}+s_{53}) (s_{61}-2 s_{62}+s_{63})\nonumber \\
&+s_{45} (s_{51}-2 s_{52}+s_{53}) (s_{61}+s_{62}+s_{63})\big) \Big) \nonumber   \\
&+2 d_1 s_{12} \Big(s_{34} \left(s_{35}-2 s_{36}\right)+s_{36}
\left(s_{35}+s_{46}\right)+\left(s_{46}-2 s_{36}\right) s_{56}\Big)+2 d_2 \Big(\left(s_{13} \left(s_{14}-2 s_{15}\right)+s_{14}
s_{15}\right) s_{26} \nonumber \\
&+\left(s_{15} \left(s_{25}-2 s_{35}\right)+s_{25} s_{35}\right) s_{46}\Big)+d_3 \Big(s_{16} \left(s_{15} \left(s_{24}-2 s_{23}\right)+s_{13} \left(s_{24}-2 s_{25}\right)+s_{14} \left(s_{23}+s_{25}\right)\right)\nonumber \\
&+\left(s_{16} \left(s_{25}-2
s_{35}\right)+s_{26} \left(s_{15}+s_{35}\right)+\left(s_{25}-2 s_{15}\right) s_{36}\right) s_{45}\Big)\nonumber  +\frac{16}{s_{123}}\Big((4c_1(s_{13}(s_{12}+s_{23})) \nonumber   \\
&+2c_2(s_{12}(s_{13}+s_{23})+s_{23}(s_{12}+s_{13})))(4c_1(s_{46}(s_{45}+s_{56}))+2c_2(s_{56}(s_{45}+s_{46})+s_{45}(s_{46}+s_{56})))\Big) \nonumber
\end{align}

\begin{align}
&+\frac{4}{3} \Bigg(\frac{1}{s_{456}}\Big(d_1 (-s_{13} s_{12}^2+s_{13}^2 s_{12}+4 s_{13} s_{23} s_{12} -s_{13}
s_{23}^2+s_{13}^2 s_{23})+d_2 (2 s_{13} s_{12}^2-2 s_{13}^2 s_{12} \nonumber   \\
&+4 s_{13} s_{23}
s_{12}+2 s_{13} s_{23}^2-2 s_{13}^2 s_{23})+d_3 (2 s_{13} s_{12}^2+2 s_{23} s_{12}^2+2 s_{23}^2 s_{12}-2
s_{13}^2 s_{23}) \Big) \Big(s_{45}-2 s_{46}+s_{56}\Big) \nonumber \\
&+\frac{1}{s_{345}}\Big(
d_1 (-s_{35} s_{34}^2+s_{35}^2 s_{34}+4 s_{35} s_{45} s_{34}\nonumber \\
&-s_{35}
s_{45}^2+s_{35}^2 s_{45})+d_2 (2 s_{35} s_{34}^2-2 s_{35}^2 s_{34}+4 s_{35} s_{45}
s_{34}+2 s_{35} s_{45}^2-2 s_{35}^2 s_{45})+d_3 (2 s_{35} s_{34}^2\nonumber \\
&+2 s_{45} s_{34}^2+2 s_{45}^2 s_{34}-2 s_{35}^2
s_{45})\Big) \Big(s_{12}+s_{61}-2 s_{62}\Big)\nonumber \\
&+\frac{1}{s_{345}}\Big(s_{34}-2 s_{35}+s_{45}\Big) \Big(d_1 (-s_{62} s_{12}^2+s_{62}^2 s_{12}+4
s_{61} s_{62} s_{12}+s_{61} s_{62}^2-s_{61}^2 s_{62})+d_2 (2 s_{62} s_{12}^2\nonumber \\
&-2 s_{62}^2 s_{12}+4 s_{61} s_{62}
s_{12}-2 s_{61} s_{62}^2+2 s_{61}^2 s_{62})+d_3 (2 s_{61} s_{12}^2+2 s_{61}^2 s_{12}-2
s_{62}^2 s_{12}+2 s_{61}^2 s_{62}))\Big)\nonumber \\
&+\frac{1}{s_{456}} \Big(s_{12}-2
s_{13}+s_{23}\Big) \Big(d_1 (-s_{46} s_{45}^2+s_{46}^2 s_{45}\nonumber \\
&+4 s_{46} s_{56} s_{45}-s_{46}
s_{56}^2+s_{46}^2 s_{56})+d_2 (2 s_{46} s_{45}^2-2 s_{46}^2 s_{45}+4 s_{46} s_{56}
s_{45}+2 s_{46} s_{56}^2-2 s_{46}^2 s_{56})\nonumber \\
&+d_3 (2 s_{46} s_{45}^2+2 s_{56} s_{45}^2+2 s_{56}^2 s_{45}-2
s_{46}^2 s_{56})\Big) \nonumber \\
&+\frac{1}{s_{561}} \Big(d_1 (-s_{24}
s_{23}^2+s_{24}^2 s_{23}+4 s_{24} s_{34} s_{23}-s_{24} s_{34}^2+s_{24}^2 s_{34})+d_2
(2 s_{24} s_{23}^2-2 s_{24}^2 s_{23}+4 s_{24} s_{34} s_{23}\nonumber \\
&+2 s_{24} s_{34}^2-2 s_{24}^2  s_{34})+d_3 (2 s_{24}
s_{23}^2+2 s_{34} s_{23}^2+2 s_{34}^2 s_{23}-2 s_{24}^2 s_{34})\Big) \Big(-2 s_{51}+s_{56}+s_{61}\Big) \nonumber \\
&+\frac{1}{s_{561}} \Big(s_{23}-2
s_{24}+s_{34}\Big) \Big(d_1 (s_{56} s_{51}^2+s_{61} s_{51}^2-s_{56}^2 s_{51}-s_{61}^2
s_{51}\nonumber \\
&+4 s_{56} s_{61} s_{51})+d_2 (-2 s_{56} s_{51}^2-2 s_{61} s_{51}^2+2 s_{56}^2
s_{51}+2 s_{61}^2 s_{51}+4 s_{56} s_{61} s_{51})+d_3 (-2 s_{61} s_{51}^2\nonumber \\
&+2 s_{56}^2
s_{51}+2 s_{56} s_{61}^2+2 s_{56}^2 s_{61})\Big)\Bigg)  +\text{cyc}(1,2,3,4,5,6) \Bigg]\Bigg\}\  ,
\end{align}
where $\text{cyc}(1,2,3,4,5,6)$ denotes cyclic permutations of $\{ 1, \dots, 6\}$, $s_{ijk} \equiv s_{ij}+s_{jk}+s_{ik}$ and $s_{ij}\equiv(p_i+ p_j)^2$.

\newpage

\section{Abelian Z-theory Amplitudes} \label{ampZ}
The Abelian Z-theory amplitudes can be found in \cite{Carrasco:2016ldy,Carrasco:2016ygv}. These amplitudes coincide with the most general color-ordered amplitudes satisfying the  KK and  BCJ relations found in \cite{Elvang:2018dco}. For completeness, we report here the results for the 4-point amplitude up to eighth derivative order, and for the 6-point amplitude up to sixth derivative order:
\begin{equation}
A_4[1,2,3,4]=\frac{C_2}{F^2}t+\frac{C_6}{F^6}t(s^2+t^2+u^2)+\frac{C_8}{F^8}t(stu)+\cdots \ ,
\end{equation}
\begin{align}
A_6[1,2,3&,4,5,6]=\frac{C_2^2}{F^4}\Bigg(\frac{s_{13}
	s_{46}}{s_{123}}+\frac{s_{15} s_{24}}{s_{234}}+\frac{s_{26} s_{35}}{s_{345}}-s_{24}-s_{26}-s_{46}\Bigg)  \nonumber \\
&+\frac{C_2C_6}{F^8}\Bigg[s_{12}^3+2 s_{234}s_{12}^2+2 s_{45} s_{12}^2-2
{s_{234}}^2 s_{12}-4 s_{23} s_{34} s_{12}+2
s_{123} s_{34} s_{12}\nonumber \\
&+4 s_{23} s_{234}
s_{12} -4 s_{123} s_{234}s_{12}+2 s_{34}
s_{234} s_{12}+2 s_{23} s_{45} s_{12}-\frac{1}{2}
s_{123} s_{45} s_{12}\nonumber \\
&+4 s_{23} s_{345}
s_{12}+s_{34} s_{345} s_{12}-\frac{1}{2} s_{45}
s_{345} s_{12}+2 s_{23} s_{56} s_{12}+\frac{1}{3} s_{34} s_{56}
s_{12} \nonumber \\
&+s_{123}
{s_{234}}^2+{s_{123}}^2
s_{234}-2 s_{23} s_{123}
s_{234}-4 s_{123} s_{34}
s_{234}+\frac{4}{3} s_{123}
s_{234}
s_{345}\nonumber \\
&-\frac{\left(s_{12}+s_{23}\right) \left(s_{12}^2+s_{23}
	s_{12}+s_{23}^2\right) \left(s_{45}+s_{56}\right)}{s_{123}}+\text{cyc}(1,2,3,4,5,6) \Bigg]+\cdots \ .
\end{align}

\section{Special Galileon Amplitudes} \label{SGA}
In this Appendix, we show the special galileon 4-point and 6-point scattering amplitudes up to $\mathcal{O}(p^{12})$. The 5-point amplitude was found to vanish up $\mathcal{O}(p^{14})$. The 4-point amplitude reads
\begin{align}
\mathcal{A}_4&=-\frac{2}{\Lambda^6}  stu+ \frac{1}{\Lambda^{10}} \left(\frac{ b_1+b_2}{40}\right)\left(s^5+t^5+u^5\right)+\frac{(c_1-6c_2)}{2^6 \Lambda^{12}}s^2t^2u^2 \nonumber \\
&\qquad \qquad \qquad \qquad +\frac{1}{\Lambda^{12}} \left(c_2+c_3+c_4\right)\left(s^6+t^6+u^6\right)+\mathcal{O}(p^{14}) \ .
\end{align}
Meanwhile, the 6-point amplitude is given by
\begin{align}
\mathcal{A}_6&= - \frac{2}{\Lambda^{12}} \big(\frac{s_{12} s_{13} s_{23} s_{46} s_{56} s_{45}}{s_{123}}+\frac{s_{12} s_{16} s_{26} s_{34}
	s_{35} s_{45}}{s_{126}}+\frac{s_{13} s_{16} s_{24} s_{25} s_{36} s_{45}}{s_{136}}+\frac{s_{14} s_{15}
	s_{23} s_{26} s_{36} s_{45}}{s_{145}}\nonumber  \\ &+\frac{s_{14} s_{15} s_{23} s_{26} s_{36}
	s_{45}}{s_{236}}+\frac{s_{13} s_{16} s_{24} s_{25} s_{36} s_{45}}{s_{245}} +\frac{s_{12} s_{16} s_{26}
	s_{34} s_{35} s_{45}}{s_{345}}+\frac{s_{12} s_{13} s_{23} s_{46} s_{56} s_{45}}{s_{456}}\nonumber  \\ &+\frac{s_{12} s_{14} s_{24} s_{35} s_{36} s_{56}}{s_{124}}+\frac{s_{12} s_{15} s_{25} s_{34} s_{36}
	s_{46}}{s_{125}}+\frac{s_{13} s_{14} s_{25} s_{26} s_{34} s_{56}}{s_{134}}+\frac{s_{13} s_{15} s_{24}
	s_{26} s_{35} s_{46}}{s_{135}}\nonumber  \\ &+\frac{s_{14} s_{16} s_{23} s_{25} s_{35} s_{46}}{s_{146}}+\frac{s_{15}s_{16} s_{23} s_{24} s_{34} s_{56}}{s_{156}}+\frac{s_{15} s_{16} s_{23} s_{24} s_{34}s_{56}}{s_{234}}+\frac{s_{14} s_{16} s_{23} s_{25} s_{35} s_{46}}{s_{235}}\nonumber  \\ &+\frac{s_{13} s_{15} s_{24}s_{26} s_{35} s_{46}}{s_{246}}+\frac{s_{13} s_{14} s_{25} s_{26} s_{34} s_{56}}{s_{256}}+\frac{s_{12}s_{15} s_{25} s_{34} s_{36} s_{46}}{s_{346}}+\frac{s_{12} s_{14} s_{24} s_{35} s_{36}	s_{56}}{s_{356}}\big)\nonumber  \\ &+\mathcal{O}(p^{14}) \ .
\end{align}

\bibliography{bibliography}

\end{document}